\documentclass{mn2e}
\usepackage{mnras_cite}
\usepackage{epsfig}
\usepackage{color}
\usepackage[bookmarks=true]{hyperref}
\usepackage{multirow}

\usepackage{graphicx}
\usepackage{subfigure}
\usepackage[T1]{fontenc}
\usepackage{calligra}

\newcommand{\rhalf}{r_{\frac{1}{2}}}
\newcommand{\dif}{{\rm d}}
\newcommand{\eq}{\begin{equation}}
\newcommand{\qe}{\end{equation}}
\newcommand{\ar}{\begin{eqnarray}}
\newcommand{\ra}{\end{eqnarray}}
\newcommand{\fig}{\begin{figure}}
\newcommand{\gif}{\end{figure}}

\begin{document}

\title[Analytic and numerical realisations of a disk galaxy]{Analytic and numerical realisations of a disk galaxy}
\author[M.~J.~Stringer, A.~M.~Brooks, A.~J.~Benson, F.~Governato]{M.~J.~Stringer$^{1,3}$, A~.M.~Brooks$^{2,3}$, A.~J.~Benson$^{2,3}$, F.~Governato$^4$ \\ 
1. Department of Physics, South Road, Durham, DH1 3LE \\
2. Theoretical Astrophysics, Caltech, 1200 E. California Blvd., Pasadena, CA 91125, U.S.A.\\ 
3. Kavli Institute of Theoretical Physics, Santa Barbara, CA 93106, U.S.A.\\
4. Department ofAstronomy, University of Washington, Box 351580, Seattle, WA98195, U.S.A.
} 
\maketitle

\begin{abstract}
Recent focus on the importance of cold, unshocked gas accretion in galaxy formation -- not explicitly included in semi-analytic studies -- motivates the following detailed comparison between two inherently different modelling techniques: direct hydrodynamical simulation and semi-analytic modelling. By analysing the physical assumptions built into the {\sc gasoline} simulation, formulae for the emergent behaviour are derived which allow immediate and accurate translation of these assumptions to the {\sc galform} semi-analytic model. The simulated halo merger history is then extracted and evolved using these equivalent equations, predicting a strikingly similar galactic system. This exercise demonstrates that it is the initial conditions and physical assumptions which are responsible for the predicted evolution, not the choice of modelling technique. On this level playing field, a {\em previously published} {\sc galform} model is applied (including additional physics such as chemical enrichment and feedback from active galactic nuclei) which leads to starkly different predictions.
\end{abstract}

\section{Introduction}

Two different approaches are widely used to test theories of galaxy formation. Both make use of developing computational resources to integrate a set of coupled differential equations forward in time, where each equation applies physical constraints, or empirical laws, to some of the system's properties.  

The semi-analytic method \cite{White91} can be characterised by the intent to encapsulate as much of the physical behaviour as possible in the equations themselves, chosen to describe {\em systematic} properties (virialisation, conservation of angular momentum, etc.) and thus leaving the minimum to numerical calculation.

Inherently different to this is the approach taken by simulations, where the ethos is to apply equations which are known to govern the {\em particular} properties (e.g. the gravitational force between two particles) and demonstrate, by skilfully programmed numerical calculation, that these reproduce macroscopic properties. 

In practice, adjustment will always need to be made to account for the fact that each simulated particle represents many, many real particles.  In traditional simulations which considered only gravity, this adjustment is purely mathematical. When gravity is the dominant interaction, results are therefore very robust and simulations are mostly accepted as being a faithful realisation of the Universe at large scales, where this is the case. A notable measure of this acceptance is that semi-analytic models have taken to using samples of halos taken directly from simulations \cite{Kauffmann99,Helly03,Hatton03}, rather than samples created statistically \cite{Bower93,Lacey93} using methods based on the analytic arguments of \scite{Press74}.

In general, though, there must come a scale below which the governing equations are no longer describing particle interactions but emergent phenomena. At these smaller scales, the two largely contrasting approaches begin to look more similar. As an immediate example, simulations which encompass an entire galaxy have not attempted to resolve individual stars and must therefore apply an equation which approximates the conversion rate of large bodies of gas into stars, based on locally averaged density.  An equivalent and sometimes identical equation appears in semi-analytic models (\S\ref{StarFormation}).

Both approaches have had some success in {\em accounting for} the observed properties of galaxies. {\em Predictive} power, on the other hand, relies on surety in the underlying equations, or a clear understanding of how differing underlying assumptions map on to final observable quantities. 

To bolster this crucial understanding, it is extremely helpful to study, in detail, the predicted evolution of a single, large galaxy and its satellites, as followed by these two different techniques. If the analytic model begins from the precise halo merger history found in the simulation, and applies the same assumptions about the physics at smaller scales, will it predict the same observable properties and, if not, where and why do the two calculations  diverge? 

\subsection{Previous research}

Existing quantitative comparisons between simulations and semi-analytic models have focussed on {\em samples} of galaxies. The approach which is adopted here -- adapting the semi analytic model to emulate the assumptions made by the simulation---was also used by \scite{Helly03} to demonstrate that the cooled mass predicted by the two approaches can be consistent. In particular, consistency was demonstrated over two orders of magnitude in halo mass. Earlier works by \scite{Benson01} and \scite{Yoshida02}, using similar techniques, reached broadly similar conclusions.

A broader comparison was made by \scite{Cattaneo07} of two predicted galaxy populations: one by an SPH model and the other by version of the GalICS semi-analytic model, modified to emulate the SPH method. Again, consistency is demonstrated between the two techniques once the semi-analytic model is appropriately ``stripped down'', but this consistency comes at the expense of agreement with observational constraints, which has been met by the full version of GalICS.

\subsection{Outline}

To extend this previous research, in this present work we will compare in detail the formation of a single galaxy in a {\sc gasoline} numerical simulation and in the {\sc galform} semi-analytic model. The structure of this paper is as follows: \\

\noindent {\bf \S 2:} A review of the broader aspects of the {\sc gasoline} code and of this particular simulation.\\

\noindent {\bf \S 3:} Detailed analysis of the way that {\sc gasoline} models each of the following key physical processes, together with an explaination of how the particular approach in question can be emulated in the {\sc galform} model.

\S3.1: Merging

\S3.2: Accretion

\S3.3: UV heating 

\S3.4: Gas distribution 

\S3.5: Cooling 

\S3.6: Disk formation 

\S3.7: Star formation 

\S3.8: Feedback 

\S3.9: Disk stability 

\noindent Particularly relevant aspects of the {\sc galform} model are reviewed (or referenced) when appropriate. General aspects of the model can be found in \scite{Cole00} and \scite{Bower06}.\\

\noindent {\bf \S 4.1:} By applying these same underlying physical assumptions to {\sc galform}, the two modelling techniques are shown to make broadly consistent predictions.\\

\noindent {\bf \S 4.2:} Having demonstrated this consistency, {\sc galform} is then used to find how this halo would evolve under the physical assumptions made in previous studies (which have been shown to successfully match the {\em collective} properties of galaxies).\\

\noindent {\bf \S 5:} Summary.

\section{GASOLINE}

The ``MW1'' simulation analyzed in this paper was generated using 
{\sc Gasoline}, an N--Body+Smoothed Particle Hydrodynamics (SPH) Parallel 
Treecode \cite{Wadsley04}, within a flat, $\Lambda$-dominated 
WMAP1 cosmology ($\Omega_0=0.3$, $\Lambda$=0.7, $h=0.7$, $\sigma_8=0.9$, 
$\Omega_{b}=0.039$).  A moderate resolution version of this run was 
originally discussed in detail in \scite{Governato07}, and the higher resolution 
version analised here has subsequently been used for a number of studies 
\cite{Zolotov09,Read09}.  High mass and  force resolution (see Table \ref{res}) was achieved using the volume renormalization technique \cite{Katz93}.  

This simulated galaxy matches the observed mass--metallicity
relationship at varying redshifts \cite{Brooks07,Maiolino08}. This
success demonstrates that these simulations overcome the historic
``overcooling problem'' \cite{Mayer08}, in which gas in simulations
cools rapidly, forming stars too quickly and early, and thus producing
a mass--metallicity relation that is too enriched at a given stellar
mass. \scite{Brooks07} showed that reproducing the mass--metallicity
relationship was a result of the adopted feedback prescription.  This
feedback mechanism, combined with high numerical resolution\footnote{With $\sim 4.8\times10^6$ total particles within the virial radius at z=0.}, avoids spurious loss of angular momentum.  As a consequence,
simulated galaxies also match the observed Tully-Fisher relationship
\cite{Mayer08,Governato08,Governato09}.

\begin{table}
\centering
\caption{Resolution details for the {\sc gasoline} simulation}\label{res}
\begin{tabular}{ll}
\hline
Dark Matter particles:&  $7.6\times 10^5M_\odot$ \\
Gas particles: & $1.3\times 10^5M_\odot$ \\
Star particles:&  $3.8\times 10^4M_\odot$ \\
Force resolution: &0.3 kpc \\
\hline
\end{tabular}
\end{table}

{\sc Gasoline}'s star formation and supernovae feedback scheme is described in detail in \scite{Stinson06}, and parameters from \scite{Governato07} are adopted for MW1.  The star formation recipe reproduces the Kennicutt-Schmidt law, and each star particle is born with a Kroupa initial mass function \cite{Kroupa93}.  Combined with a prescription for a cosmic UV background  \cite{Haardt96} that mimics reionization and unbinds non-collapsed baryons from simulated halos with total masses below about $10^9 M_{\odot}$, this feedback scheme drastically reduces the number of luminous satellites containing a significant mass of stars, avoiding the well known ``substructure problem'' \cite{White91,Kauffmann93,Quinn96,Moore99}.

As described in \scite{Stinson06}, supernova energy is deposited into the ISM mimicking the blast-wave phase of a supernova (following the Sedov-Taylor solution).  Cooling is turned off for gas particles within the supernova
blastwave radius for the duration that the remnant is expected to adiabatically expand (\S\ref{Feedback}).  Supernova feedback regulates star formation efficiency as a function of halo mass,  resulting in the mass--metallicity relationship described above.  This regulation of star formation also leads to realistic trends in gas 
fractions, with the lowest mass galaxies being the most gas rich \cite{Brooks07}, and reproducing the observed incidence rate and column densities of Dampled Lyman $\alpha$ systems at $z =$ 3 \cite{Pontzen08}.

MW1 and its most massive progenitor halo were identified at each simulation output step using AHF\footnote{{\bf{A}}MIGA's {\bf{H}}alo {\bf{F}}inder, available for download at \href{http://www.aip.de/People/aknebe/AMIGA}{\tt http://popia.ft.uam.es/AMIGA}} \cite{Gill04,Knollmann09}.  AHF determines the virial radius, $r_{\rm v}$, for each halo at each output time step using the overdensity criterion for a flat universe following \scite{Gross97}.  The full gas accretion history of MW1 is described in detail in \scite{Brooks09}, and used in this study for comparison to {\sc Galform}.

\section{Translation from simulation to semi-analytics}

\subsection{The merger tree}\label{MergerTree}
The common point between the two different models of galaxy formation is the halo merger tree. This is generated by finding bound structures in the ``MW1'' simulation of \scite{Brooks09} using AHF. A virial mass is assigned to each structure and its fate at each timestep is categorised, being either: 
\begin{itemize}
\item[1.]{Isolated;} 
\item[2.]{Inside the virial radius of a larger system; or,}
\item[3.]{Merged.}
\end{itemize}
Two criteria are used to classify a galaxy as ``merged'':
\begin{itemize}
\item[(a)]{If the DM mass decreases by more than 50\% in a single timestep; or,}
\item[(b)]{If the number of DM particles falls below 64.} 
\end{itemize}
These are both significantly different from the definition adopted by {\sc galform} (see \pcite{Cole00}), so a detailed comparison of merger times, along the lines of \cite{Jiang08}, is deferred for another study.

When a structure changes from category 1 to category 2, the {\sc galform} code updates its status to a ``satellite'' and predicts its subsequent evolution based on its approach velocity, {\bf v} and position {\bf r} at this juncture. These two properties are combined into the dimensionless parameter $\Theta$ (\S\ref{ThetaOrbit}), which characterises the orbit. The distribution of values for $\Theta$ in this simulation was found to be consistent with the most recent assumption applied in {\sc galform} (Fig. \ref{Theta}).

This orbital information is then used to estimate a merger time based on the standard Chandrasekhar formula (\ref{t_mrg}) as defined by \scite{Lacey93}. Details of this part of the model can be found in Appendix~\ref{Satellites} and in \ncite{Cole00}~(\gmcite{Cole00}; \S 4.3).

The resulting dark and baryonic components are shown in Fig. \ref{infall}. Being generated purely from the derived merger tree, the two realisations might be expected to be identical. One reason this is not quite the case is because the total virial mass, $M_{\rm v}$, is measured directly, but the relative composition of baryons and dark matter may vary a little from halo to halo, and from one realisation to the other.

\fig
\includegraphics[trim = 62mm 154mm 10mm 50mm, clip, width=\columnwidth]{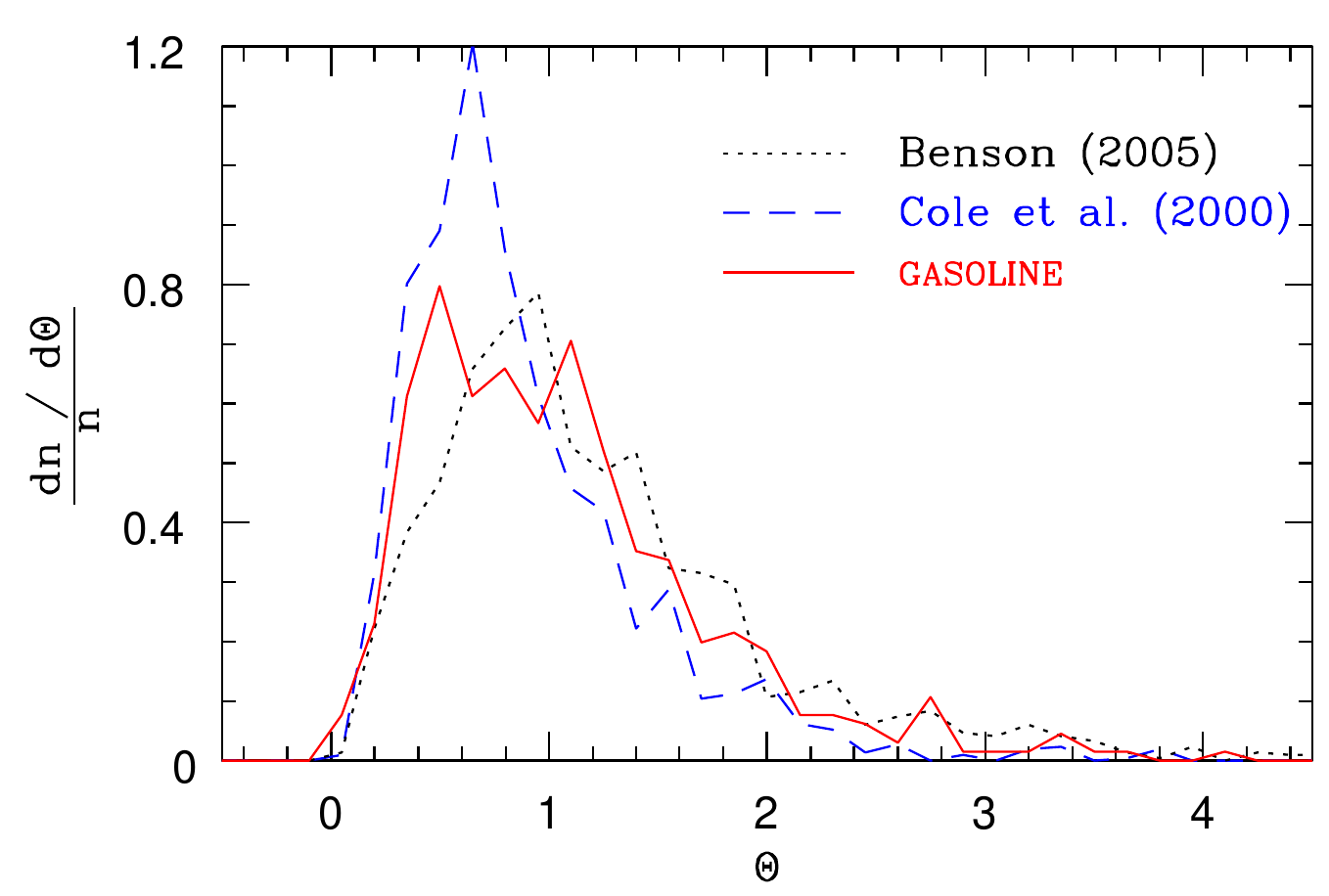}
\caption{Orbital parameters randomly selected from the distribution (\ref{OrbitDist}) of \protect\scite{Cole00} are shown in the dashed line, while the dotted line shows the distribution from \protect\scite{Benson05}. The solid line shows the distribution found in the simulation of \protect\scite{Brooks09} (which results from applying equation (\ref{OrbitFactor}) to the positions and velocities of the satellites). }\label{Theta}
\gif

Previously, {\sc galform} required that the halo mass was monotonically increasing as a function of time along each branch of the merger tree (i.e. each halo should be more massive than the sum of the masses of its progenitor halos). While this condition is always fulfilled for merger trees generated using the extended Press-Schechter theory, it is not necessarily fulfilled for merger trees extracted from N-body simulations.

For this reason, previous use of N-body merger trees in {\sc galform} has required some adjustment of halo masses\footnote{The standard approach which was taken begins at the earliest times in the tree, checks whether each halo is more massive than the sum of the masses of its progenitor halos and, if it is not, sets its mass to be very slightly larger than that sum. This process was repeated for all halos in the tree.} to enforce this monotonic behaviour \cite{Helly03}. Unfortunately, this adjustment is somewhat arbitrary and can lead to
significant changes in halo mass when adjustments are propagated through the merger tree.

Monotonicity in halo masses along branches of the merger tree is not enforced in this work. Instead, they are allowed to follow whatever evolution the N-body simulation provides. This requires some modification of the way in which
{\sc galform} assigns baryonic masses to halos. Previously, each halo was assigned an initial mass of hot gas equal to:
\eq
M_{\rm hot}=\frac{\Omega_{\rm b} }{\Omega_{\rm M} } \left[M_{\rm v}({\rm new})-\sum_{\rm progenitors} M_{\rm v}({\rm progenitor})\right]~,\label{mhot}
\qe
where $M_{\rm v}$ was the total mass of the halo and ``progenitors'' refers to halos that exist on a previous timestep but have, at some point, merged into the current ``parent'' (or its most massive progenitor). If halo masses do not monotonically increase, this could lead to negative gas masses. To ensure against this, the initial mass of gas is assigned to be the greater of (\ref{mhot}) and zero.

The results are identical to the original implementation in cases where the halo masses are monotonically increasing, but this can result in the total baryonic mass not being precisely equal to the fraction $(\Omega_{\rm b}/\Omega_{\rm M})$ of the total mass of the final halo. Fortunately, this difference is found to be small and, in any case, the simulated halos do not have precisely this ratio either.

\fig
\includegraphics[trim = 67mm 170mm 10mm 52mm, clip, width=\columnwidth]{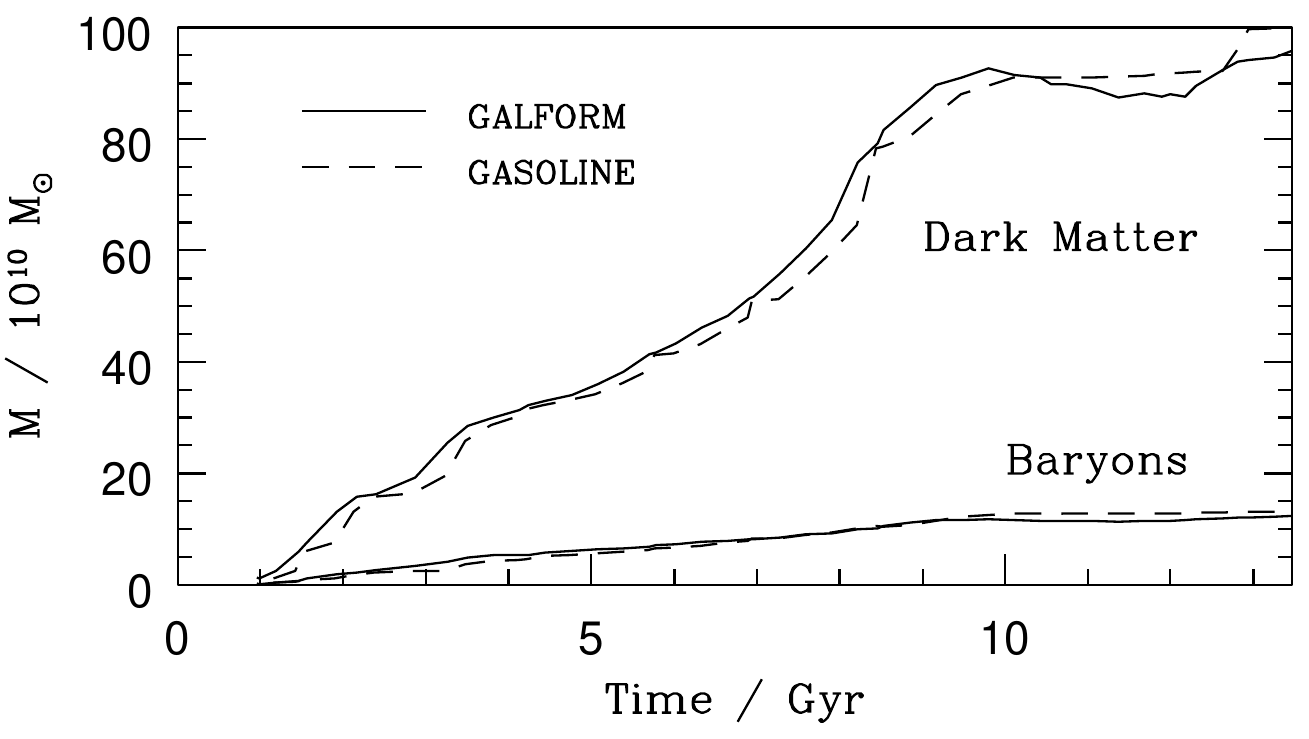}
\caption{Confirmation that the mass evolution of the ``MW1'' system which emerged in the {\sc gasoline} simulation of \protect\scite{Brooks09} (solid lines) matches that contained in the extracted merger history used by {\sc galform} (dashed lines).}\label{infall}
\gif

\subsection{Accretion}\label{Accretion}

The temperature history of accreted gas was investegated by \scite{Keres05} in a simulation containing 1120 galaxies. The distribution of gas particles in terms of their maximum past temperature is bimodal, suggesting that classification into hot and cold modes of accretion is appropriate. The contribution from each of these modes is then found to depend strongly on halo mass and on environment. Notably, the halo mass ($M_{\rm v}\sim3\times10^{11}M_\odot$) which separates low-mass  galaxies dominated by cold accretion and higher mass galaxies dominated by the hot mode, is close to that found by analytic arguments \cite{Birnboim03}.

The decomposition of the accreting gas into different modes is a principle feature of the analysis by \scite{Brooks09}. The three modes of accretion are defined as follows (along with their analogues in the semi-analytic realisation):

\begin{itemize}
\item[(a)]{{\bf Clumpy acccretion}} is defined to include ``any particles that have, at any previous output step, belonged to a galaxy halo other than the main halo we are considering.''  \cite{Brooks09}

The information needed to evaulate this component in the semi-analytic model is readily available from the merger tree, being the virial masses, $M_{\rm v}$, of the existing, progenitor halos:
\eq\label{clumpy}
M_{\rm clumpy}({\rm new}) = \frac{\Omega_{\rm b}}{\Omega_{\rm M}}\sum_{\rm progenitors}M_{\rm v}({\rm progenitor})~.
\qe
The subsequent fate of any baryons within these substructures is decided by their inbound trajectory and considerations of dynamical friction, as described in \S\ref{MergerTree} and Appendix~\ref{Satellites}.\\
\fig
\includegraphics[trim = 67mm 55mm 10mm 52mm, clip, width=\columnwidth]{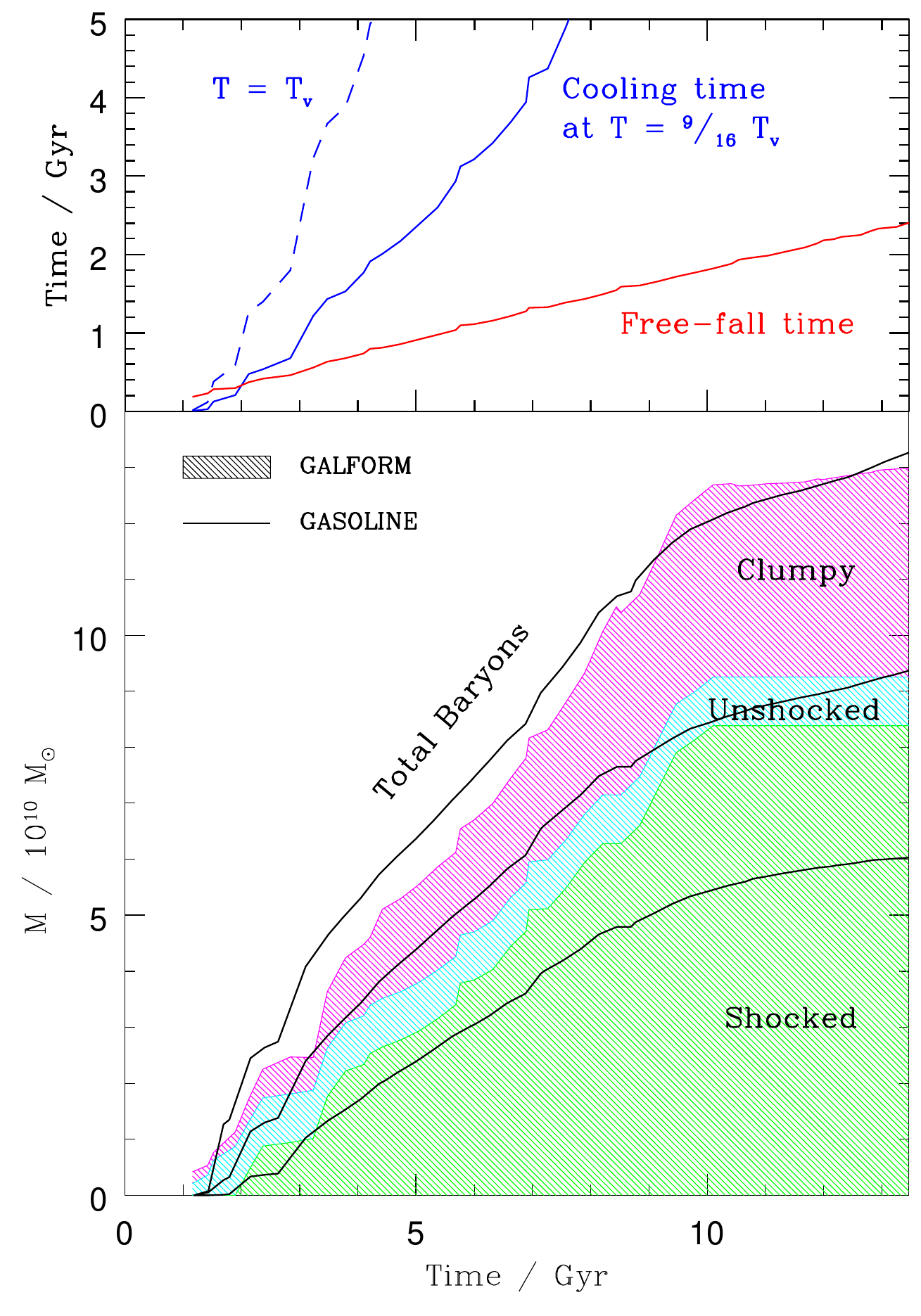}
\caption{An analysis of accretion onto the MW1 galaxy in {\sc gasoline}, in the light of assumptions associated with {\sc galform}. The top panel shows the cooling time for a gas particle at the virial radius (\ref{t_cool}) at two relevant temperatures, together with the timescale for freefall under gravity (\ref{t_ff}).  In the lower panel, the solid lines show how accretion is divided amongst these three modes in the simulation. The shaded areas show how the equivalent components would be estimated in {\sc galform}. As a simple attempt to emulate (\ref{ShockConditions}a), shocked and unshocked are distinguished by comparing the cooling time at $T=~^9/_{16}T_{\rm v}$ to the the freefall time (as explained in \S\ref{Accretion}b). } \label{accretion}
\gif

\item[(b)]{{\bf Smooth accretion: Shocked component}}

The following criteria are required for a gas particle in the simulation to be categorised as shocked:
\eq
(a)~T\ge \frac{3}{8}\mathcal{T}_{\rm v} \hspace{0.6cm}{\rm and}\hspace{0.6cm} (b)~\Delta\rho > 4\left(\frac{3\mathcal{T}_{\rm v}}{3T}\right)^\frac{3}{2}-1~,\label{ShockConditions}
\qe
where $\mathcal{T}_{\rm v}$ is the virial temperature defined in {\sc gasoline}
\eq
k_{\rm B}\mathcal{T}_{\rm v}\equiv\frac{GM_{\rm v}\mu m_{\rm H}}{3 r_{\rm v}}~.\label{T_v}
\qe
and $\Delta\rho$ is the fractional increase in density since the previous timestep. The latter is an entropy criterion and the factor of 4 is motivated by the Rankine-Huginot shock jump conditions for a singular isothermal sphere. Though low resolution can artificially broaden shocks, \scite{Brooks09} verify that this run is of sufficiently high
resolution to resolve shock discontinuities.

The virial temperature in {\sc galform} is defined differently, being a factor of $3/2$ higher, $T_{\rm v}\equiv\frac{3}{2}\mathcal{T}_{\rm v}$, so (\ref{ShockConditions}a) corresponds to $T \ge\frac{9}{16}T_{\rm v}$.

Finding an appropriate analogue to this condition within the framework of {\sc galform} is difficult. However, one relevant calulation that {\em is} made compares the cooling timescale for gas at temperature $T$:
\eq
t_{\rm cool}(r) = \frac{3\mu m_{\rm H}k_{\rm B}T}{2\rho_{\rm gas}(r)\Lambda(T, Z)}~,\label{t_cool}
\qe
and the free fall time from a radius, $r$, enclosing mean density $\bar{\rho}$:
\eq
t_{\rm ff}(r) = \sqrt{\frac{3}{8\pi G\bar{\rho}(r)}}=\frac{r_{\rm v}}{2}\sqrt{\frac{\mu m_{\rm H}}{k_{\rm B}T_{\rm v}}}\label{t_ff}
\qe
The cooling function $\Lambda$ is found for given temperature $T$ and metallicity, $Z$, by reference to the published tables of \scite{Sutherland93}. $\rho_{\rm gas}$ is the gas density, $\bar{\rho}(r)$ the mean density of all the mass enclosed by radius $r$, and $\mu m_{\rm H}$ is the mean particle mass.

So, in the context of {\sc galform} accreting gas is designated as shocked if:
\eq
t_{\rm cool}(\mathcal{T}_{\rm v},r_{\rm v})>t_{\rm ff}(\mathcal{T}_{\rm v},r_{\rm v})~,\label{GalformShocked}
\qe
this being a reasonable, simple equivalent to (\ref{ShockConditions}a).\\

\item[(c)]{{\bf Smooth accretion: Unshocked component}}

Gas in the simulation which fails to meet either of the criteria in (\ref{ShockConditions}) is categorised as unshocked. Similarly, accreting gas in {\sc galform} is designated as unshocked if $t_{\rm cool}<t_{\rm ff}$.\\ 
\end{itemize}

The relative contributions of these three components are shown in Fig. \ref{accretion} for both realisations.  Being determined directly from the merger tree, it is unsurprising that the clumpy components are quite consistent. 

That the shocked component is overestimated by the semi-analytic model is entirely as anticipated by \scite{Brooks09}, though the eventual impact this has on the predicted properties of the galaxy may not be quite as severe, as will be seen in the forthcoming sections.

\subsubsection{Alternative criterion}

\scite{Birnboim03} investigate the conditions for the existence of virial shocks in galactic halos, culminating in the following expression for the existence of a shock:
\eq
\frac{\rho_{\rm gas}(r_{\rm v})r_{\rm v}\Lambda(T_{\rm v})}{4\left(\mu m_{\rm H}\right)^2\left|u_0\right|^3} < 0.0126~. \hspace{1.5cm}\left[u_0^2 = \frac{16k_{\rm B}T_{\rm v}}{3\mu m_{\rm H}}\right]\label{BD1}
\qe
Referring to equations (\ref{T_v}), (\ref{t_cool}) and (\ref{t_ff}) above, this condition can be re-written:
\ar
t_{\rm ff}&<&0.2~t_{\rm cool}~.\label{BD}
\ra
As can be seen from Fig.\ref{accretion}, the cooling time does not exceed the free-fall time by a factor of 5 until about 8 Gyr. The criterion (\ref{BD}) would therefore imply that all smoothly accreted gas will be unshocked before that point, corresponding to an unshocked:shocked ratio of 2 or 3, as opposed to $1/2$ in gasoline or $1/6$ according to (\ref{GalformShocked}).

So, for this particular mass aggregation history, one simple estimate (\ref{GalformShocked}), overestimates the shocked component in the simulation and another, more thorough estimate (\ref{BD1}), gives an underestimate. This suggests that the error in assuming spherical symmetry may unfortunately exceed any accuracy that can be gained by analysis of shock conditions. Calibrating some of these analytic models to simulations may help quantitatively account for the complex geometry.

\subsection{UV heating}

A UV background is incorporated by \scite{Brooks09} which is based on an updated model by \scite{Haardt96}. This background ``turns on at $z=6$'' and heats the gas in halos so that only those with ``~100 or more dark matter particles are of sufficient mass to retain bound gas particles''. For the resolution in this case, 100 DM particles corresponds to $M_{\rm cut} = 7.6\times 10^7M_\odot$. 

The {\sc galform} model is build with two corresponding parameters: A redshift, $z_{\rm cut}$, below which the UV background heating is effective (set to $z_{\rm cut}=6$ for consistency with the above) and a halo velocity, $v_{\rm cut}$ below which all cooling is suppressed. For a virial overdensity, $\Delta_{\rm v}=300$, the mass cut-off, $M_{\rm cut}$, mentioned above translates to a cut-off halo velocity of:
\eq
v_{\rm cut}= 
\left(\Delta_{\rm v}/2\right)^\frac{1}{6}\left(1+z\right)^\frac{1}{2}\left(\Omega_{\rm M}{\rm G}M_{\rm cut}{\rm H}_0\right)^{\frac{1}{3}}\approx 30~{\rm km~s}^{-1}
\qe
These values are duly applied, but one key difference between the modelling of this process remains. The UV background in the simulation can {\em slow} cooling in larger halos without necessarily completely suppressing it, while in {\sc galform} suppression is all or nothing (below and above $v_{\rm cut}$ respectively).

\begin{figure*}
\includegraphics[trim = 7mm 55mm 10mm 50mm, clip, width=\textwidth]{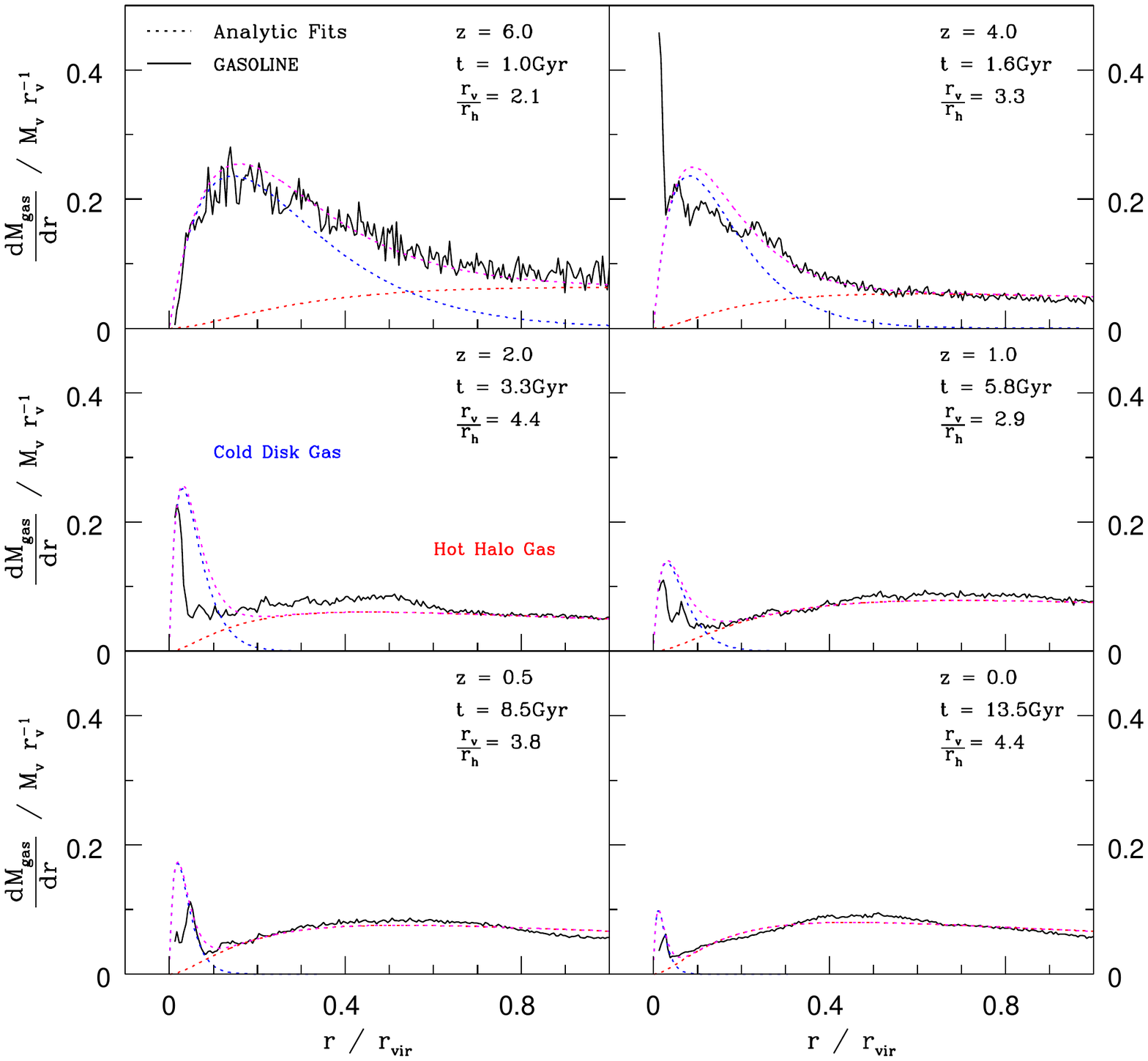}
\caption{The total gas density profile from the simulation of \protect\scite{Brooks09} at six different redshifts, shown as solid lines.  The dotted lines show the analytic forms which are assumed by the {\sc galform} model for the disk gas (eqn. \ref{RadialProfile}, blue line) and diffuse halo gas (eq. \ref{rho_halo}, red line). The sum total of these two components is also shown (magenta line). Each of these depend on a single parameter, fitted respectively by the disk scalelength, $r_{\rm D}$ and the core radius of the hot gas profile, $r_{\rm H}$. (The integral under each analytic component is already fixed to match the total hot and cold gas mass within the virial radius.) Because there are so many more points to fit the hot component, these are much better matches than the cold gas. As it is the hot scalelength ({\em not} the cold gas scalelength) which is actually an input to {\sc galform}, this convenient bias has not been corrected.}\label{halogas}
\end{figure*}

\subsection{Hot gas distribution}\label{HotGas}

The distribution of hot gas that is assumed in the {\sc galform} model was recently updated, in the light of work by \scite{Sharma05}, to the following density profile for hot gas of mass $M_{\rm H}$:
\eq\label{rho_halo}
\rho(r) = \frac{\rho(0)}{\left(1+ r/r_{\rm H}\right)^3}~.~~~\left[\rho(0)\equiv \frac{ M_{\rm H}/4\pi r_{\rm H}^3 }{ \int_0^{r_{\rm v}/r_{\rm H}}x^2(1+x)^{-3}\dif x}\right]
\qe
This form was assumed in recent work on disk formation and structure \cite{Stringer07},  and will also be used here.  

The reason for this choice is clear from Fig. \ref{halogas}, which demonstrates that it is a satisfactory description of the distribution of diffuse gas in the outer parts of this simulated system. In this Figure, the analytic form for the hot halo gas (\ref{rho_halo}) and cold disk gas (\ref{RadialProfile}) have been fitted to the simulation results, yielding the respective scalelengths:
\begin{itemize}
\item{Halo core radius, $r_{\rm H}$.}
\item{Disk scalelength, $r_{\rm D}$.}
\end{itemize}
(with the initial constraint on each line that it integrates to the correct total mass for that component.) These scalelengths, found from fitting the simulation results, can be scrutinised against those deduced from the usual analytic assumptions.

Firstly, the halo core radius is typically assumed to be a constant fraction of the virial radius, which is not quite the case in this particular simulation (Fig. \ref{halogas}). However, in the absence of any obvious pattern in the evolution of $r_{\rm H}$, the usual simple scaling is applied, with the choice: 
\eq\label{r_core}
r_{\rm H} = r_{\rm v}/3~,
\qe
being the best approximation. Fig. \ref{Radii} compares this to values found from fitting to the true gas distributions in Fig. \ref{halogas}, the result looking favourable. This straightforward scaling with the virial radius (\ref{r_core}) will therefore be used in the matched {\sc galform} realisation. The cold gas scalelengths, also shown in Fig.\ref{Radii}, are not inputs to {\sc galform}, but are calculated within the model as described in \S\ref{DiskFormation}.

To understand the relevance of the hot gas distribution on the system's evolution as a whole, it is worth reviewing the precise treatment of gas cooling in {\sc galform}. 

\fig
\includegraphics[trim = 67mm 55mm 10mm 135mm, clip, width=\columnwidth]{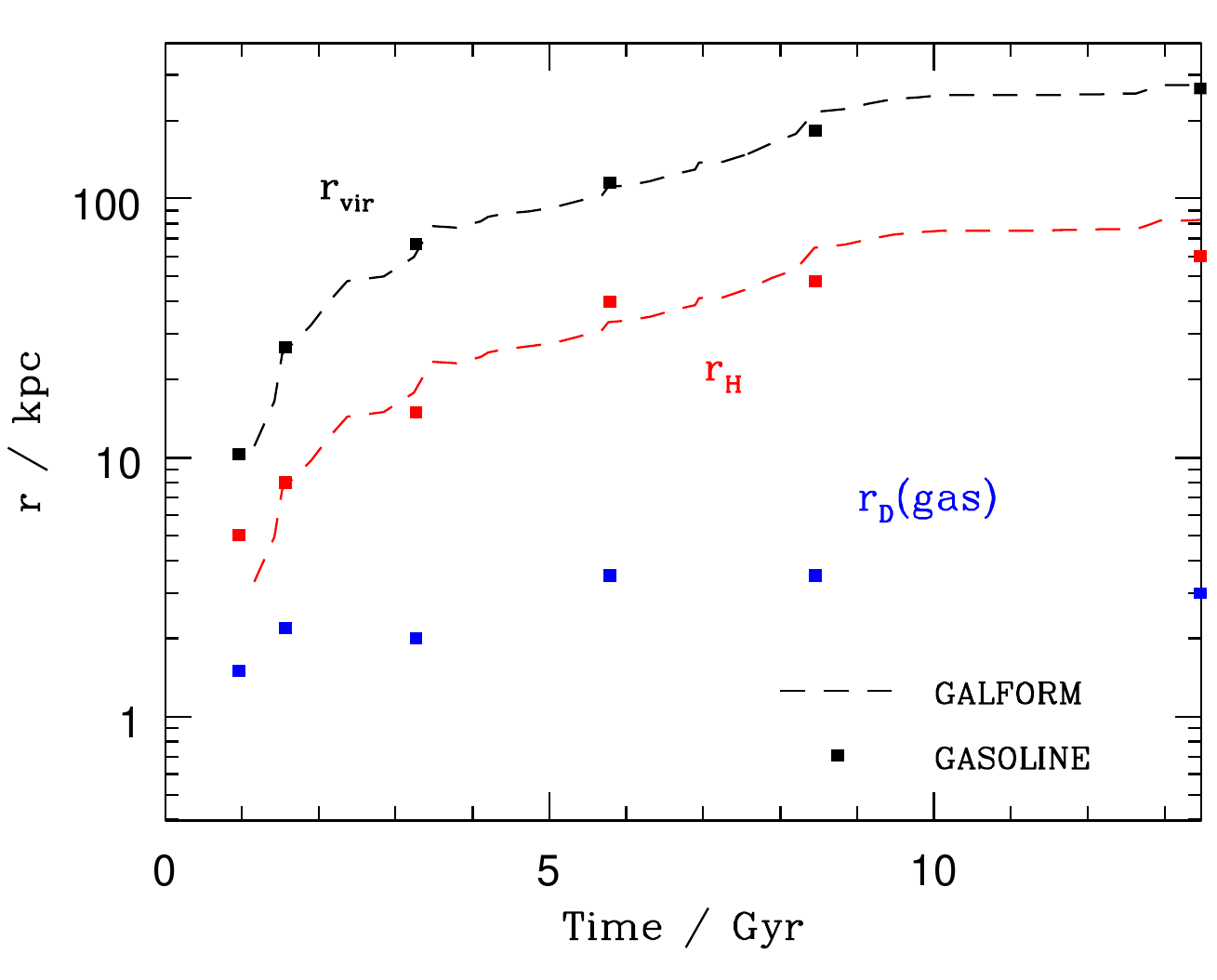}
\caption{The evolution of three key scalelenghts of the system: Virial radius, $r_{\rm v}$, halo gas scalelength, $r_{\rm H}$ (\ref{rho_halo}), and disk scalelength, $r_{\rm D}$ (\ref{RadialProfile}). Dashed lines show the values that will be assumed by {\sc galform}. Points show the values which best fit the gas profile of the simulation (Fig. \ref{halogas}).}\label{Radii}
\gif

\subsection{Cooling}\label{Cooling}

The cooling model described by \scite{Cole00} determines the 
mass of gas able to cool in any timestep by following the propagation of the 
cooling radius in a notional hot gas density profile\footnote{We refer to this 
as a ``notional'' profile since it is taken to represent the profile before any 
cooling can occur. Once some cooling occurs presumably the actual profile 
adjusts in some way to respond to this and so will no longer look like the 
notional profile, even outside of the cooling radius.}. This profile is fixed when a 
halo is flagged as ``forming'' and is only updated when the halo undergoes 
another formation event. The mass of gas able to cool in any given timestep is 
equal to the mass of gas in this notional profile between the cooling radius at 
the present step and that at the previous step. The cooling time is assumed to 
be the time since the formation event of the halo. Any gas which is reheated 
into or accreted by the halo is ignored until the next formation event, at 
which point it is added to the hot gas profile of the newly formed halo. The 
notional profile is constructed using the properties (e.g. scale radius, virial 
temperature etc.) of the halo at the formation event and retains a fixed 
metallicity throughout, corresponding to the metallicity of the hot gas in the 
halo at the formation event.

This work makes use of a new cooling model, which will be described in full detail in \scite{Benson09}. Rather than arbitrary ``formation'' events, this model uses a continuously updating estimate of cooling time and  halo properties. The properties described in \S\ref{HotGas} (density normalization, core radius) are  reset at each timestep. The previous infall radius\footnote{defined as the lesser of the cooling and freefall radii.} (i.e. the radius within which gas was allowed to infall and accrete onto the galaxy) is computed by finding the radius which encloses the mass previously removed from the hot component in the current notional profile.

The new model must supply an alternative estimate of the time available for cooling in the halo, $t_{\rm avail}$,  from which the cooling radius can be computed in the usual way (i.e. by finding the radius in the notional profile at which $t_{\rm cool}=t_{\rm avail}$). This is done by considering the energy radiation rate {\em per particle}, $\dot{\epsilon}_{\rm r}$ and the thermal energy {\em per particle}, $\epsilon_{\rm th}$, which are estimated by making standard assumptions:
\eq
\epsilon_{\rm th}=\frac{3}{2}k_{\rm B}T_{\rm v}\hspace{1cm}{\rm and}\hspace{1cm}\dot{\epsilon}_{\rm r}=\Lambda(T_{\rm v},Z)n~.\label{e_th}
\qe
In terms of these quantities, the cooling time (\ref{t_cool}) is simply:
\eq
t_{\rm cool}(r,t) = \frac{\epsilon_{\rm th}(t)}{\dot{\epsilon}_{\rm r}(r_{\rm cool}, t)}~.  \label{tcool}
\qe
At any time, the model needs to identify some radius in the hot halo, $r_{\rm cool}$,  where the gas has had just enough time to radiate all its thermal energy. This {\em as yet undetermined} radius is defined to satisfy the condition that the cooling time for gas particles at this point is equal to the time available for them to cool, which is estimated in terms of the mean energy radiation rate per particle\footnote{This is proportional to the mean number density $\bar{n}$:
\eq
\overline{\dot{\epsilon}_{\rm r}}\equiv\frac{\Lambda\int_0^Nn(N')\dif N'}{N}=\alpha(r_{\rm H})\Lambda\bar{n}
\qe
The factor $\alpha$ does depend on halo structure (i.e. on $r_{\rm H}$) but, if the distribution of hot gas {\em relative to the virial radius} is the same for each halo, as implied in this case by (\ref{r_core}), then this factor cancels out in (\ref{fudge2}).}, $\overline{\dot{\epsilon}}_{\rm r}$:
\eq
t_{\rm avail}(t) = \frac{\int_0^t\overline{\dot{\epsilon}_{\rm r}}(t_i)\dif t_i}{\overline{\dot{\epsilon}_{\rm r}}(t)}~. \label{fudge}
\qe
Equating (\ref{tcool}) and (\ref{fudge}) then gives:
\ar
t_{\rm cool} &=& t_{\rm avail} \nonumber \\
\frac{\frac{3}{2}k_{\rm B}T_{\rm v}}{\Lambda(t)n(r_{\rm cool},t)} &=& \frac{\int_0^t\Lambda(t_i)\overline{n}(t_i)\dif t_i}{\Lambda(t)\overline{n}(t)}~,\label{fudge2}
\ra
where $\bar{n}$ is the mean number density inside the virial radius. 

This approach must account for the hierarchical assembly of a halo, so the denominator in the right hand side strictly includes a sum over all progenitors, $i$, of this system which exist at time $t_i$. Incorporating this sum into (\ref{fudge2}), and rearranging, then yields a final expression to be solved for the cut-off density, $\rho_{\rm cool} \equiv \rho(r_{\rm cool})$:
\eq
\frac{\rho_{\rm cool}(t)}{\bar{\rho}(t)} = \frac{\frac{3}{2}k_{\rm B}T_{\rm v}}{\int_0^t\left[\sum_i\Lambda_i(t_i)\bar{n}_i(t_i)\right]\dif t_i}~.\label{r_cool}
\qe

This equation is notably independent of halo structure, a factor which only comes in when the limiting density is used to compute the total cooled mass. The scalelength, $r_{\rm H}$, appears in the normalisation of the hot gas profile (\ref{rho_halo}), and hence effects the mass {\em enclosed} by the cut-off density, $\rho_{\rm cool}$.
\fig
\includegraphics[trim = 64mm 55mm 10mm 162mm, clip, width=\columnwidth]{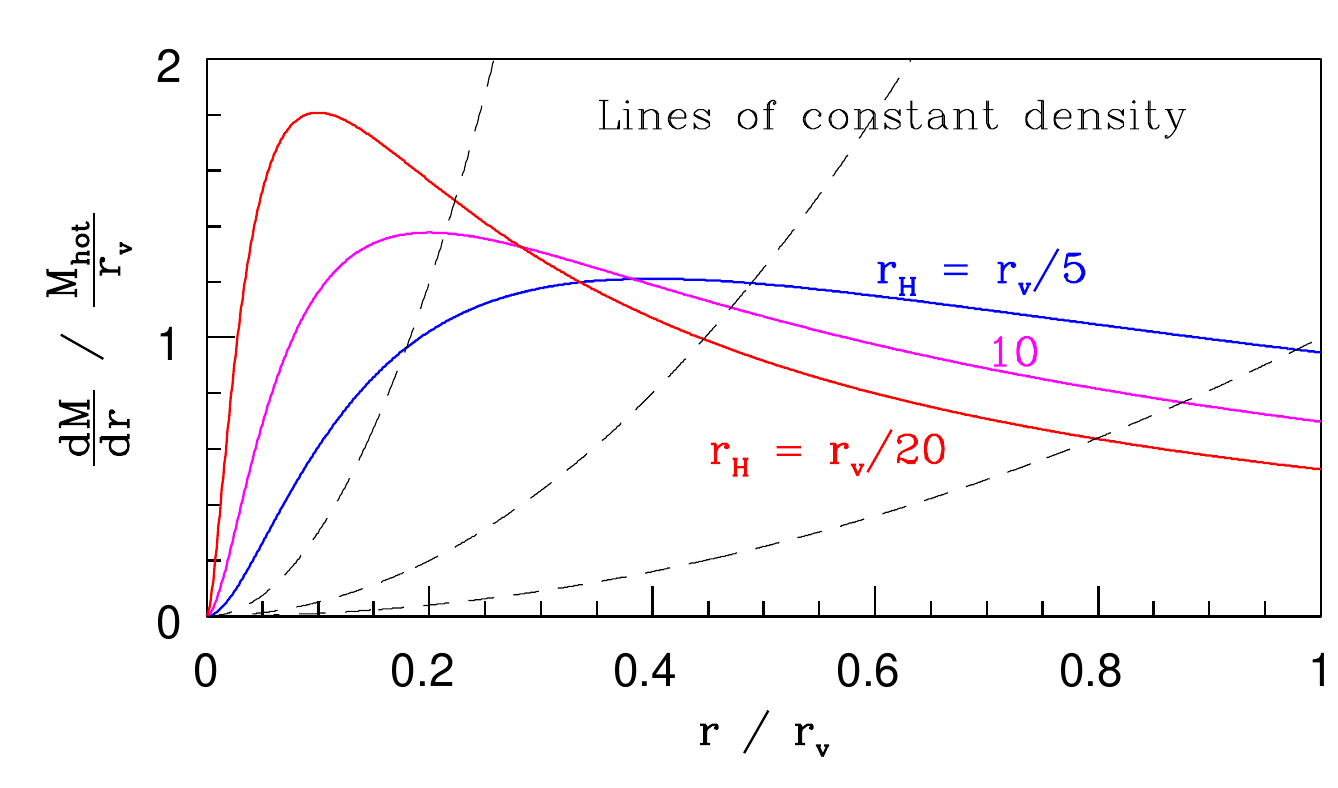}
\caption{Three normalised hot gas distributions (solid lines), corresponding to three different values of the scalelength $r_{\rm H}$ applied to the profile (\ref{rho_halo}). Intersections between a solid line and one of the dotted lines correspond to solutions to equation (\ref{r_cool}) for a particular value of $r_{\rm H}$ and a particular cut-off density. The cooled fraction, $M_{\rm cool}/M_{\rm hot}$, is given by the integral under the solid curve up to the intersection with the relevant cut-off density.}\label{haloprofile}
\gif

This can be appreciated graphically. Solutions to (\ref{r_cool}) are illustrated in Fig. \ref{haloprofile} for three choices of $r_{\rm H}$ and three examples of cut-off density. For the higher cut-off density, a smaller $r_{\rm H}$ (more centrally concentrated) will lead to a greater cooled mass. For the case of a low cut-off density, a centrally concentrated profile will actually lead to a lesser cooled mass, contrary, perhaps, t

\subsection{Disk formation}\label{DiskFormation}

In {\sc galform}, galactic disks are assumed to form from the cooled gas, conserving the net angular momentum that it possessed when it was part of the halo. The assumed distribution for the specific angular momentum, $j$, of the hot halo gas is taken from \scite{Sharma05}:
\begin{equation}
\frac{\dif M(j)}{\dif j} \propto j^{\alpha-1}\exp\left(-\frac{\alpha j}{\langle j\rangle}\right)~.\label{haloAM}
\end{equation}
The value of $\alpha$ was found to lie between 0.5 and 1.5, but the mean value $\alpha=0.89$ is adopted here. This is normalised to enclose the correct total mass and the mean specific angular momentum of the hot gas, $\langle j\rangle$, which is found directly from the spin parameter,
\eq
\lambda\equiv \frac{\langle j\rangle\left |E\right|^{1/2}}{{\rm G}M_{\rm v}^{3/2}}~.\hspace{1.3cm}\left[\langle j\rangle=\frac{1}{M_{\rm v}}\int_0^{\rm M_{\rm v}}j\dif M(j)\right]\label{lambda}
\qe
Details of the calculation of the halo energy, $E$, can be found in \ncite{Cole00}~(\gmcite{Cole00}; \S3.2.1). The disk scalelength, $r_{\rm D}$, is recalculated after each timestep such that angular momentum is conserved in the system. The calculation is based on the following surface density distribution (applied to each disk component, $i$): 
\eq
\Sigma_i(R)  = \Sigma_i(0)e^{-\frac{r}{r_{\rm D}}}~,\hspace{1.2cm}\left[{\rm so}~~M_i=2\pi\Sigma_i(0)r_{\rm D}^2\right]\label{RadialProfile}
\qe
and a circular speed, $v_{\rm c}$, which is constant with radius. The subsequent evolution of the disk is then followed by applying physical assumptions which are as close as possible to those made in the simulation, as is explained in the following sections.

As with the sub-halo trajectories, the parameter $\lambda$ can ordinarily be assigned at random, from a specified distribution. In this case, its value is measured for each of the simulated halos and passed on to {\sc galform}, along with the rest of the merger tree information (\S\ref{MergerTree}).

The goal here is not to prove that parameters like $\lambda$, and $\Theta$ must necesarily be set by simulations in order to validate the results of any semi-analytic model. Rather, it is an example of the correspondence between these initial conditions and the properties of the galaxy which they produce. This should reassure theorists that, when such models are applied to large samples of galaxies, their individual properties will be representative as long as these governing, random, but {\em physical} parameters are drawn from the correct natural distributions.

\subsection{Star formation}\label{StarFormation}

At face value, {\sc gasoline} and {\sc galform} would appear to be applying the same assumptions regarding the conversion of cold gas to stars. These respective assumptions can be found from equation (4) of \scite{Stinson06} and equations (4.4) and (4.14) of \scite{Cole00}\footnote{An additional factor involving the circular speed appears in (4.11) of \scite{Cole00}. This is omitted here (equivalent to setting $\alpha_\star=0$).}: 
\eq\label{sfrs}
(a)~~\frac{\dif\rho_\star}{\dif t} = c_\star\frac{\rho_{\rm cold}}{t_{\rm dyn}}\hspace{1cm}{\rm and}\hspace{1cm} (b)~~\psi = \epsilon_\star\frac{M_{\rm cold}}{\tau_{\rm disk}}~.
\qe
where $\psi$ is the total instantanoues star formation rate in the whole disk, and $\rho_\star$ is the local value per unit volume. The timescales which appear in these expressions have similar physical motivation, but are not identical. One, $t_{\rm dyn}$, is the time for gravitational collapse of a spherically symmetric region of local density $\rho_{\rm cold}$. The other, $\tau_{\rm disk}$, is the angular period of the whole system at the half-mass radius, $\rhalf$:
\eq
(a)~~t_{\rm dyn} \equiv \frac{1}{\sqrt{G\rho_{\rm cold}}}\hspace{0.8cm}{\rm and}\hspace{1cm}(b)~~\tau_{\rm disk} \equiv\frac{r_{1/2}}{v_{\rm c}}~.
\qe
So, despite the apparent similarity of the two expressions in (\ref{sfrs}), they can lead to quite different instantaneous {\em total} star formation rates, $\psi$. To illustrate this, one can consider the surface density distribution (\ref{RadialProfile}) which {\sc galform} assumes, together with some decrease in density away from the plane of the disk:
\eq\label{Scaleheight}
\rho_i(R,z) = \rho_i(R,0){\rm e}^{-\frac{\left|z\right|}{h}}~.\hspace{0.8cm}\left[{\rm so}~~\Sigma_i(R) = 2\rho_i(R,0)h\right]
\qe
Applying the local star formation assumption of (\ref{sfrs}a) to this distribution gives:
\ar
\psi_a 
&=& 2\pi\sqrt{G} \rho_{\rm cold}(0,0)^{\frac{3}{2}}\int_{-\infty}^\infty e^{-\frac{3\left|z\right|}{2h}}\dif z\int_0^\infty e^{-\frac{3r}{2r_{\rm D}}}R~\dif R \nonumber \\
&=& c_\star \left(\frac{2}{3}\right)^3M_{\rm cold}^{3/2} \sqrt{ \frac{G}{4\pi hr_{\rm D}^2} }~.\label{CorrectedSFR}
\ra
Comparison between this and the usual prescription in {\sc galform} exposes the big {\em physical} difference between the  two star formation models; (\ref{sfrs}a) considers the gravitational collapse of the gas due to its own gravity while (\ref{sfrs}b) includes the gravity of all components\footnote{The dark matter does influence the creation of stars due to its effect on the overall stability, but this is a separate part of the analytic treatment (\S\ref{DiskStability}).}, including the stars and dark matter ($M_{\rm tot}$ rather than just $M_{\rm cold}$).

The derived analytic equivalent (\ref{CorrectedSFR}) is shown in Fig.\ref{SFtimescale} (using the fitted values of $r_{\rm D}$ from \S\ref{HotGas} and the illustrative approximation\footnote{This choice is not too critical since $h$ appears only to the one half power in eqn. \ref{CorrectedSFR}).} of a constant scaleheight, $h=200$pc) alongside the full simulation results.  Discrepancies are larger at early times when the assumed cold gas gas density distribution is tenuous, but the match improves as the disk settles and is excellent at late times, showing that (\ref{CorrectedSFR}) is as consistent with the assumptions in {\sc Gasoline} as is possible within this particular analytic framework.

\subsubsection{Burst star formation}

During merger events, and if the disk is deemed to be unstable (\S\ref{DiskStability}), cold disk gas in the {\sc galform} model is assumed to funnel to the centre of the new system and be converted to stars within a timestep, which in this case is about 0.2Gyr (\ncite{Cole00}~\gmcite{Cole00}; \S4.3). The incorporation of this non-equilibrium effect should broadly correct  the total star formation rate on occasions when the assumed density profile (\ref{RadialProfile}) is not such a good description of the true gas distribution. This is relevant here in the early stages of the simulation (as apparent when comparing the points and solid line in Fig. \ref{SFtimescale}). 

\subsubsection{Schmidt-Kennicutt Law}

Equation (\ref{sfrs}b) can be recognised as a version of the established empirical relation between star formation rate and gas supply \cite{Kennicutt98}, for which the constant of proportionality has the value $\epsilon_\star\approx0.017$. The corresponding star formation efficiency for this simulated system can be calculated from its approximate half-mass radius and circular velocity (Fig. \ref{rotation}) and this is shown in Fig. \ref{SFtimescale} as a guide.

\fig
\includegraphics[trim = 62mm 171mm 10mm 53mm, clip, width=\columnwidth]{disk}
\caption{The star formation efficiency from the simulation is shown as a solid line. The points show the result of applying the corrected star formation rate (\protect\ref{CorrectedSFR}) to the disk scalelengths, $r_{\rm D}$ which best fit the simulation (Fig.\ref{halogas}). The illustrative approximation of a constant scaleheight, $h=200$pc, is used. The feint dotted line shows the efficiency found in empirical studies, which is given by eqn. (\ref{sfrs}b) with $\epsilon_\star=0.017$ \protect\cite{Kennicutt98}.}\label{SFtimescale}
\gif
\subsection{Feedback}\label{Feedback}

\subsubsection{Heating of disk gas}
The MW1 simulation treats each type II supernova event as a blast wave which expands to a maximum radius, $r_{\rm max}$, and the enclosed volume is assumed to be unable to cool for as long as the shell can be expected to survive, $t_{\rm max}$. If the star formation rate varies little over this timescale, and the shell subsequently cools rapidly, this will correspond  to a heated mass,
\eq
M_{\rm heat}(t) \approx \frac{\dot{M}_\star t_{\rm max}}{m_{\rm SN}}\left(\frac{4\pi\rho_{\rm cold}r_{\rm max}^3}{3}\right)\equiv\dot{M}_\star(t)t_{\rm heat}~,\label{mSN}
\qe
where $m_{\rm SN}$ is the average mass of stars formed per supernova. Now, (\ref{mSN}) conceals the fact that both $r_{\rm max}$ and $t_{\rm max}$ (and hence $t_{\rm heat}$) are, rightly, functions of the local gas density, $\rho$ and pressure, $P$. The precise dependence assumed is found in equations 9 and 11 of \scite{Stinson06}. Combining this into (\ref{mSN}) gives:
\eq
t_{\rm heat} \sim \rho (\rho^{-0.16}P^{-0.2})^3\rho^{0.34}P^{-0.7}\sim\rho^{0.86}P^{-1.3}~.\label{scaling}
\qe

Referring to the phase diagram in figure 4 of \scite{Stinson06}, gas in star forming regions appears to be approximately isothermal at $T\approx 10^4$K (which is expected due to the steep gradient in the cooling curve at this temperature). This leads to the suggestion that the environmental dependence in (\ref{scaling}) will be minimal, the pressure and density terms almost cancelling for constant temperature. Though tenuous, this provides at least a rough explaination for the extremely simple dependence which does indeed emerge from the full calculation, as shown in Fig. \ref{feedback}:
\eq
M_{\rm heat}(t) \approx \dot{M}_\star(t)t_{\rm heat}~.\hspace{1.5cm}\left[t_{\rm heat}\approx 7\times 10^6 {\rm yrs}\right]\label{t_heat}
\qe
The precise definition of $M_{\rm heat}$ used in the figure is the mass of disk gas which fails to satisfy the temperature  criterion for star formation ($T<30,000$K). Thus, the integrated effect of feedback will be to slightly reduce the fraction of the total disk gas which is available for star formation\footnote{For a massive galaxy such as this, the star-forming timescale is typically $\tau_\star>10^8$ years, so the correction to allow for re-heating actually makes very little difference.  Quick examination of the inset to Fig. \ref{feedback} confirms this; the fraction of the total gas which gets heated above the star-forming limit is tiny.}. 
\eq
\dot{M}_\star = \frac{M_{\rm cold} - M_{\rm heat}}{\tau_\star}  = \frac{ M_{\rm cold} }{ \tau_\star+t_{\rm heat} }
\qe
where $\tau_\star$ is the particular timescale derived from equation (\ref{CorrectedSFR}). This revision is duly applied in {\sc galform}, being a fair reflection of the physical assumptions in the simulation.

\fig
\includegraphics[trim = 62mm 154mm 10mm 53mm, clip, width=\columnwidth]{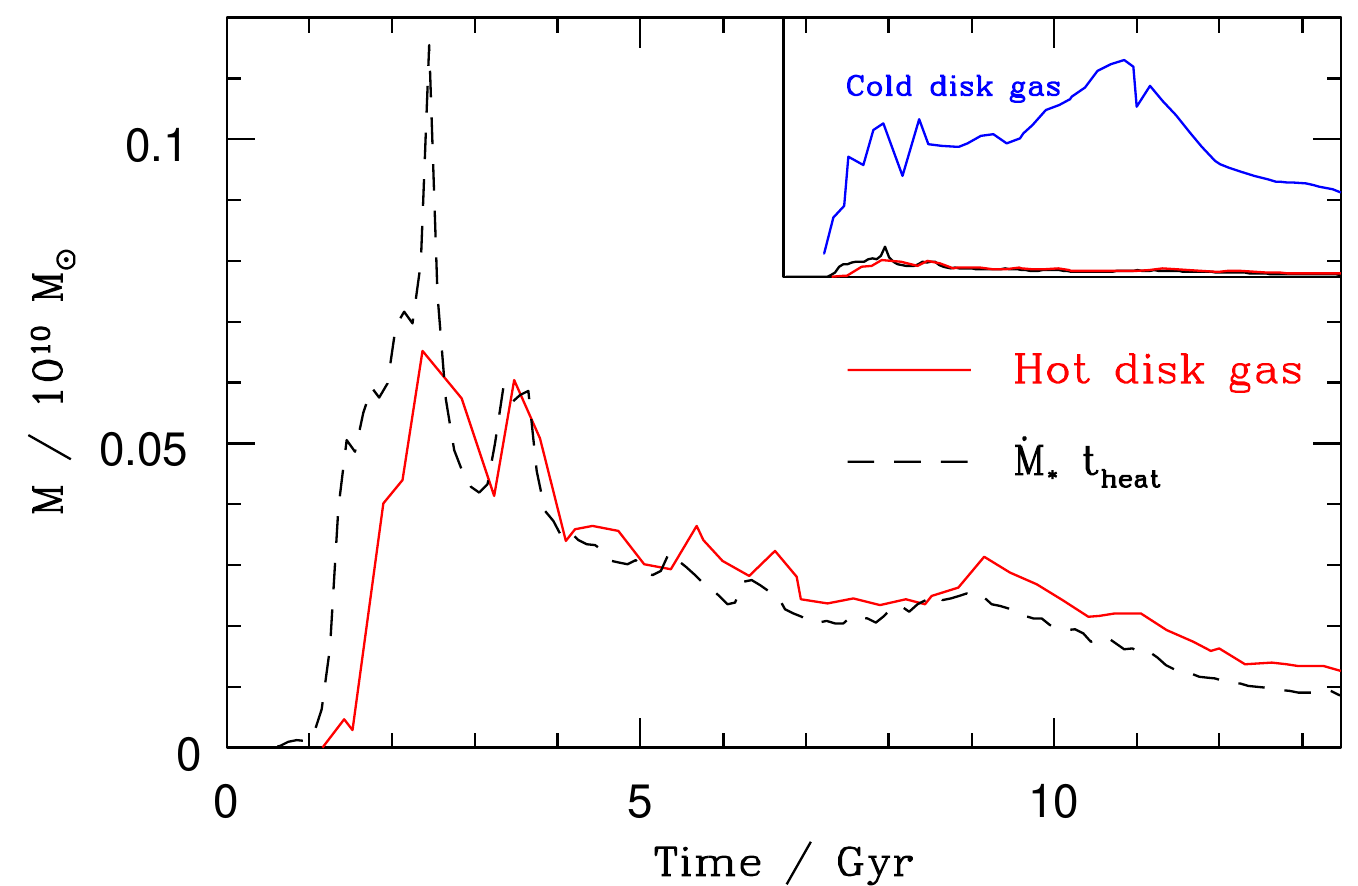}
\caption{A simple illustration of the correlation between star formation rate and the mass of heated gas present in the disk. The effective timescale from equation (\ref{t_heat}), $t_{\rm heat}=7\times 10^6$ years.  The inset panel gives an indication of the almost negligible fraction of the total disk gas which this represents. }\label{feedback}
\gif

\subsubsection{Outflow}
To discover the extent of outflow from the galaxy in the simulation, all particles were identified which have moved from the inside to the outside of a radius of 30 comoving kpc.  The total mass of such material turned out to be even less than the heated disk mass (see Fig. \ref{feedback}) and can safely be ignored for the purposes of this study.  The outflow of gas as a result of supernovae in the central galaxy is therefor set to zero in this version of {\sc galform}, for consistency with the {\sc Gasoline} simulation. 

It is important to note that this feedback model is more effective at lower halo masses, due to shallower potential wells \cite{Brooks07}. This dependence is also accounted for in the usual {\sc galform} feedback model, where the rate of mass outflow is assumed to be at least equal to the star formation rate, and thousands of times higher for smaller systems \cite{Bower06}. The consequence of this difference is investigated in \S\ref{BowerModel}. 

\subsection{Disk stability}\label{DiskStability}
\fig
\includegraphics[trim = 62mm 170mm 10mm 53mm, clip, width=\columnwidth]{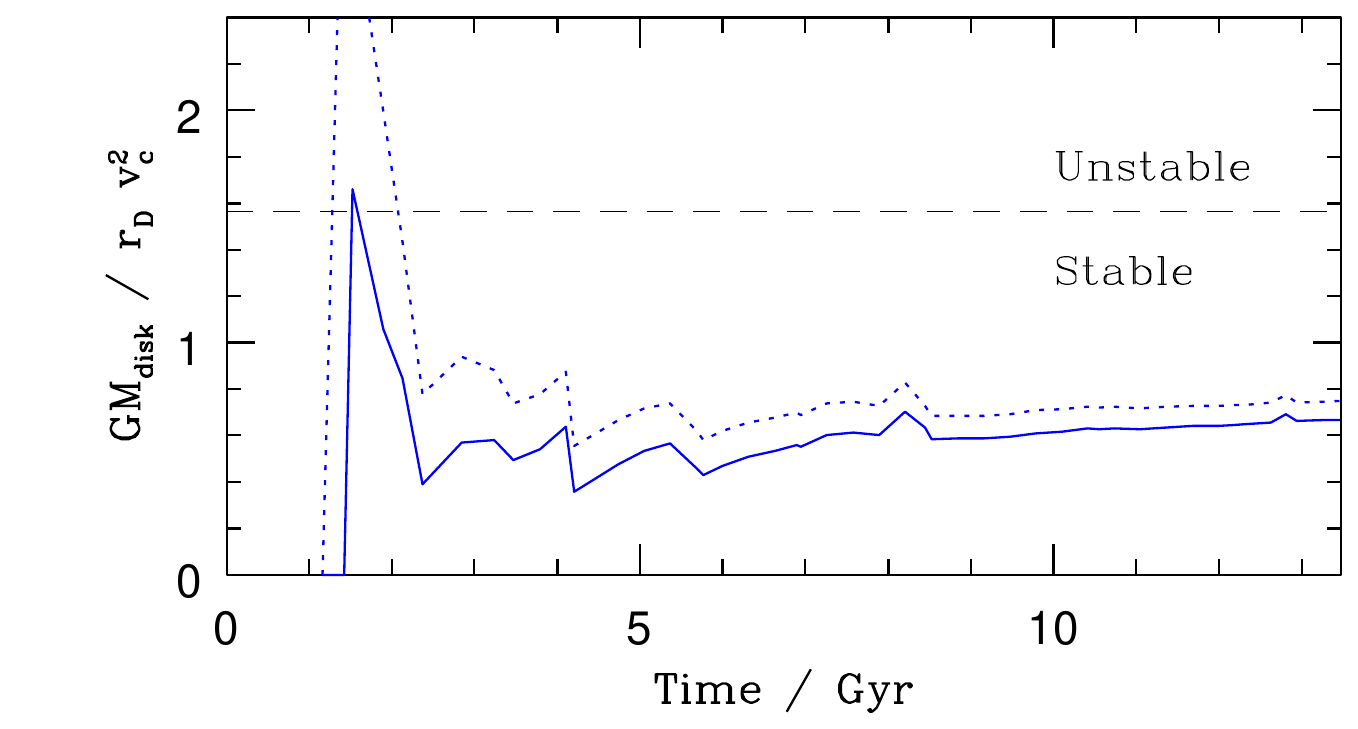}
\caption{The dimesionless ratio which is used as a stability critereon in the {\sc galform} model, shown for this particular realisation (solid line) and for a relisation where the collapse of ``unstable'' disks has not been enforced (dotted line).}\label{stab}
\gif
The analytic condition which is used to assess the stability of disks in the {\sc galform} model follows \scite{Efstathiou82}. It deems disks to be unstable to bar formation if:
\eq
 \frac{GM_{\rm disk}/r_{\rm D}}{v_{\rm c}^2}> 1.56~.\label{StabilityLimit}
\qe
When this inequality is met, all disk material is transferred to a central bulge. 

In the usual {\sc galform} model, some set fraction of disk mass is also transferred to a central black hole \cite{Bower06}. This was used to arrive at the value in (\ref{StabilityLimit}) which was found to best match the Magorrian relation between bulge mass and black hole mass, $M_{\rm BH} \sim M_{\rm bulge}^{1.12}$, observed by \scite{Haring04}.  In this study, the black hole accretion fraction is set to zero for the purposes of consitency with the simulation, where this aspect of evolution is not considered.
\fig
\includegraphics[trim = 67mm 115mm 10mm 53mm, clip, width=\columnwidth]{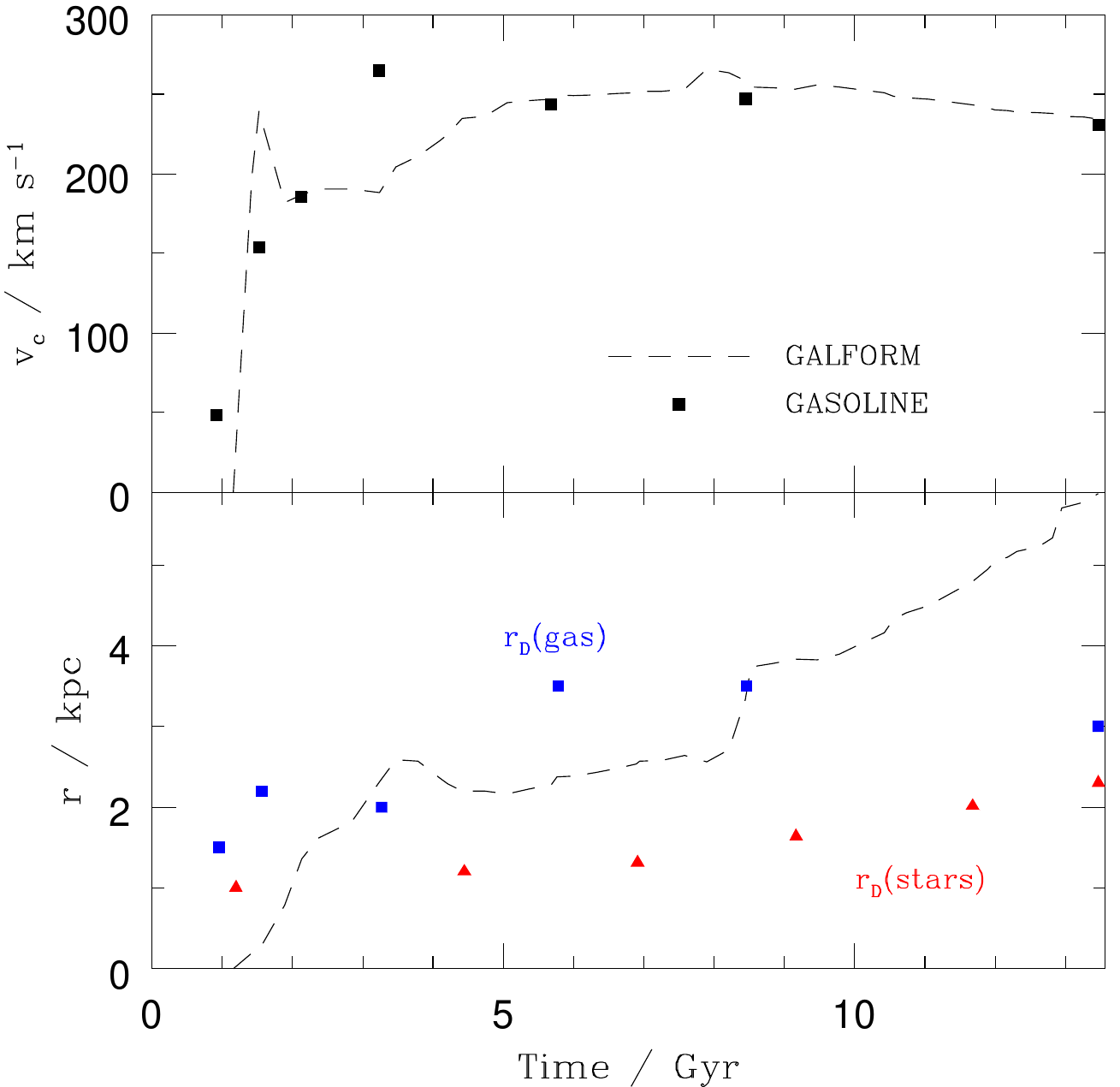}
\caption{The evolution of the disk rotation velocity, $v_c$, (top panel) and the disk scalelength, $r_{\rm D}$, (lower panel) for both realisations. The circular velocity of the simulated disk (black squares) is taken at 5 comoving kpc from the centre. The approximate gas scalelengths (blue squares) are found by fitting (\ref{RadialProfile}) to the simulated gas distributions in Fig. \ref{halogas}. The stellar scalelengths (red triangles) are taken from fitting (\ref{RadialProfile}) the the stellar profiles, shown later in Fig. \ref{stars}. Dashed lines are the values calculated in {\sc galform} from conservation of angular momentum (\S\protect\ref{DiskFormation} - gas and stars are assumed to have the same scalelength).}\label{rotation}
\gif
Being physically well motivated, the condition (\ref{StabilityLimit}) has been applied throughout this study. Conveniently, the precise formalism used need not cause too much concern here; the change that its application produces in the component masses (in Fig. \ref{baryons}, for example) is barely enough to be noticed. This can be understood from Fig. \ref{stab}, which shows that only a very limited adjustment to the disk mass is required to prevent the limit in (\ref{StabilityLimit}) from being reached. The mass aggregation of this particular system, and the fact that the specific  angular momentum of the gas is relatively high\footnote{
The criterion  (\ref{StabilityLimit}) can also be written:
\eq
M_{\rm disk} > 1.2[ M_{\rm DM}(r_{1/2}) + M_{\rm sph}(r_{1/2})]~,
\qe
where $M_{\rm DM}$ and $M_{\rm sph}$ are the masses of the dark matter and spheroid components. The disk's eventual half mass radius will enclose more dark matter if the original spin parameter is higher.} ($\lambda > 0.07$ for the latter half of the halo's history) leads to the stabilisation of the system, even if the redistribution of mass is not enforced.

\section{Results}\label{Results}

\subsection{Matched Model}

\fig
\includegraphics[trim = 67mm 55mm 10mm 125mm, clip, width=\columnwidth]{infall}
\caption{The evolution of hot halo gas, cold disk gas and stars according to both modeling approaches. The star formation rate in {\sc galform} is calculated using assumptions identical to those in the simulation (\ref{sfrs}a), but taking into account their integrated effect over the whole disk profile (\ref{RadialProfile}).}\label{baryons}
\gif

\subsubsection{Structure}

The size and rotation speed are shown for both realisations in Fig. \ref{rotation}. The predicted {\sc galform} value is based on conserving the initial angular momentum (\ref{haloAM}) of in-falling gas; an assumption which is particularly effective in this case where the distributions for the initial and final mass were appropriate (Fig.\ref{halogas}). This correspondence is encouraging for the approach of calculating disk sizes from the spin parameter, $\lambda$, of the halo from which the gas cooled.

\subsubsection{Cooled Mass}

The total cooled mass (stars and cold gas), calculated using the methods of \S\ref{Accretion}, can be seen from Fig. \ref{baryons}. Agreement with the simulation is excellent, given the inherent difference in the methods of calculation, and bolsters confidence in the rather intricate calculation of the energy radiated by the system throughout its complex merger history (\ref{e_th}-\ref{r_cool}).

The division of the cooled mass into stars and gas are also shown in Fig. \ref{baryons}. The evident agreement can be understood by reference to Fig's \ref{halogas}, \ref{SFtimescale} and \ref{feedback}, which demonstrate that the considerable complexities of the simulation reduce to relatively simple relationships when integrated over the entire disk.

\subsubsection{Discrepancies}

There is a minor discrepancy at early times ($t<8$Gyr) in the cooled mass generated by the two modelling techniques. This is unlikely to be due to differences in cooling from the hot halo: given the similar density distributions (Fig. \ref{halogas}) and cooling function, the two models should yield similar amounts of cooled gas, all else being equal.

The discrepancy is also not due to a difference in the time of development of a hot gas halo.  The top panel of Fig. 4 shows that accreted gas in the {\sc galform} model transitions from unshocked, cold gas to shocked, hot gas at an age of about 2 Gyr, while the bottom panel shows that this time is also when the {\sc gasoline} model begins to develop a hot gas halo.  

This demonstrates that the adopted analytic separation of unshocked and shocked gas (free fall limited regime versus cooling limited regime) is an excellent analog to the transition between cold gas accretion and shocked gas accretion seen in simulations as a galaxy crosses the threshold mass capable of developing a hot halo. It is important to note that this transition has always been included in analytic models of galaxy formation, and thus cold gas accretion at low galaxy mass does not alter the standard picture of galaxy formation.  

The discrepancy in cooled mass arises after the development of this hot halo. The simulated galaxy continues to accrete cold gas via filaments that penetrate within the hot gas halo  (see Fig. \ref{accretion} here, figure 5 of \scite{Brooks09} and also \scite{Ocvirk08}, \scite{Agertz09} and \scite{Dekel09}), while cold gas accretion ends in the analytic model.

\subsection{Adapting {\sc gasoline} to emulate semi-analytics}
\fig
\includegraphics[trim = 67mm 115mm 10mm 53mm, clip, width=\columnwidth]{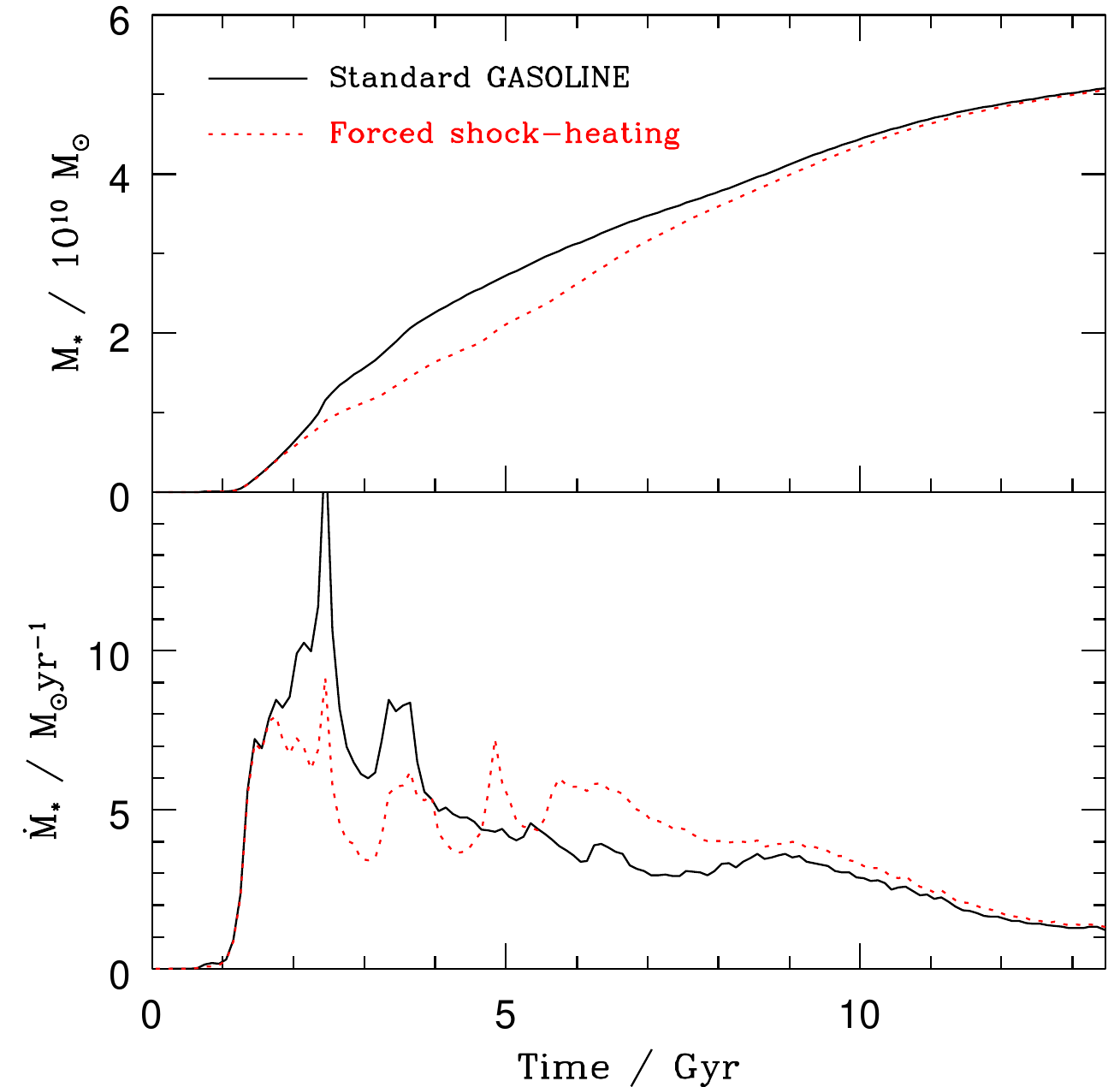}
\caption{Having focussed on modifications to {\sc galform}, this figure shows the effects of modifying the SPH code {\sc gasoline} to emulate the semi-analytic technique. The solid black line is the star formation rate (lower panel) and its integral (upper panel)  which occurs in the standard {\sc gasoline} simulation which has been under discussion. The dotted red lines show the consequence of delaying star formation from cold accreted gas is delayed by a cooling time. }\label{DelayedGrowth}
\gif

\scite{Brooks09} demonstrated that the inclusion of cold gas accretion along filaments leads to an earlier phase of star formation than predicted if all gas was initially shock heated, due to the shorter cooling times onto the central galaxy of the cold gas.  Fig. 12 shows that the simulated galaxy does indeed form stars earlier than the analytic model, though the overall discrepancy is never more than a factor of two.

In an attempt to adapt the {\it simulations} to include similar physics as the {\it analytic} model, Fig. \ref{DelayedGrowth} shows the effect on the stellar growth in the simulations if all of the cold gas had instead been shocked.  The hot gas has cooling times several
Gyr longer than the cold gas.  This is quantified, and a delay has been added to the formation time of the stars spawned from
the cold accreted gas \cite{Brooks09}.  As seen in Fig. \ref{DelayedGrowth}, this delay leads to a slower build of up stellar
mass in the simulations. Comparison with Fig.\ref{baryons} shows this to be in even better agreement with the stellar growth of the analytic model, which neglects this cold gas accretion along filaments.

\subsection{Comparison with standard {\sc galform} assumptions}\label{BowerModel}

In order to represent the same physical assumptions as this particular simulation, the semi-analytic code had to be significantly modified from the version which has been used most recently to study the collective properties of galaxy samples \cite{Bower06,Stringer09}. There are three very significant differences between this version of {\sc galform} and the simulation studied here, all of which are manifest in the comparison of the history of the simulation and the \scite{Bower06} model (Fig. \ref{Bower}).
\fig
\includegraphics[trim = 67mm 55mm 10mm 52mm, clip, width=\columnwidth]{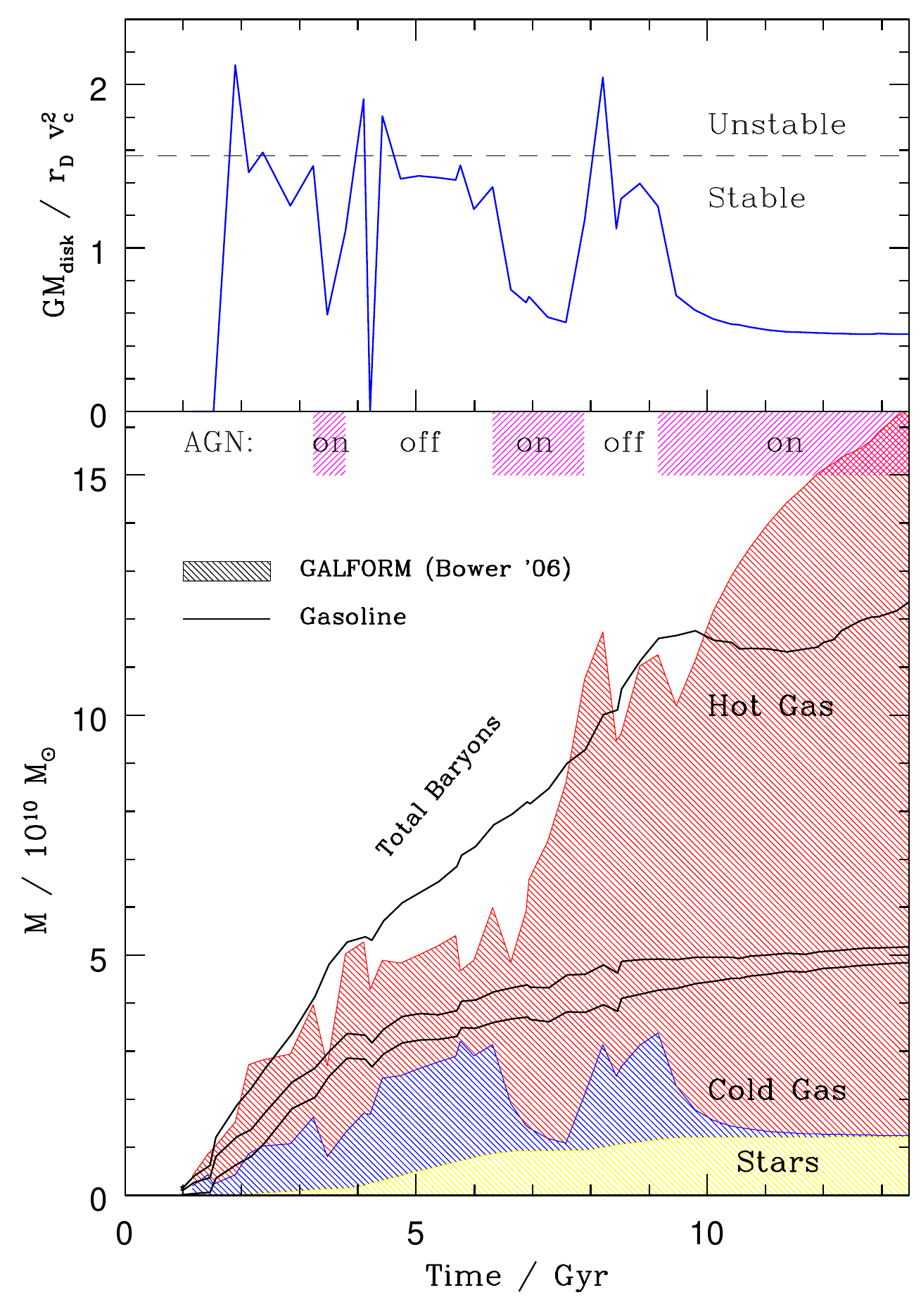}
\caption{The results of applying the unadjusted version of {\sc galform} \protect\cite{Bower06} to the same merger tree as used for Fig. \protect\ref{baryons} and  indeed the whole paper. To bring out the contrast between the two models, the shading now represents the {\sc galform} predictions and the lines show the simulation results. The upper panel is the equivalent to Fig. \protect\ref{stab}, showing the stability of the disk in this version for this model.}\label{Bower}
\gif

\subsubsection{Inclusion of AGN feedback}

When there are instabilities (\S\ref{DiskStability}) or merger events in this model which trigger disk collapse, 5\% of the available disk gas is assumed to accrete onto a central black hole. If the following criteria are satisfied, it is assumed that no hot halo gas will be able to cool onto the disk. 
\begin{itemize}
\item[(1)]{The resulting Eddington luminosity of the black hole is at least 4\% of the cooling luminosity of the halo.}
\item[(2)]{The cooling time (\ref{t_cool}) exceeds the freefall time (\ref{t_ff}) by some chosen factor (in this case, $t_{\rm cool}>0.58t_{\rm ff}$).}
\end{itemize}
This effect is visible in Fig. \ref{Bower} as a period of rapidly decreasing cold gas mass; during such phases it is no longer being replenished by gas cooling in from the halo.

\subsubsection{Stronger supernova feedback}

This is assumed to be extremely strong in the \scite{Bower06} model, the justification being that a more modest conversion of supernova energy to gas outflow would allow the formation of too many low-mass galaxies. (Indeed, it appears that the predicted number may still be too high \cite{Stringer09} even with this generous allocation of feedback energy.)

In conjunction with the inclusion of AGN heating, this assumption leads to a system with a much greater hot gas component: As soon as the supply of gas to the disk can no longer be replenished by cooling, all the gas in the system quickly ends up in the hot component. This effect is additionally enforced by the low star formation efficiency (see below), which prevents disk mass being locked up into stars.

\subsubsection{Inefficient star formation}

The star formation assumptions in {\sc Gasoline} and the \scite{Cole00} model have already been contrasted in \S\ref{StarFormation}. The assumed efficiency in the \scite{Bower06} model is to be so low ($\epsilon_\star=0.0029$) that, even after allowing for the structure and gas fraction, the final conversion rate is about a factor of 10 {\em lower} than in the simulation. ($\dot{M}_\star/M_{\rm cold}\approx$ 0 to 0.2/Gyr as opposed to the rate of 0.5 to 3/Gyr seen in Fig. \ref{SFtimescale}).

This stands out in Fig \ref{Bower}. The large mass of cold gas which is supported by the disk in the semi-analytic realisation (except when cooling is shut off by AGN) cannot be sustained in the simulation due to much more effective conversion to stars. 

\section{Summary}\label{Summary}

The merger history of a simulated, Milky Way-like galaxy has been used to recreate the same system using the {\sc galform} semi-analytic model. 

The incoming trajectories of satellite halos were investigated and found to be consistent with the usual, assumed distributions. Information on the different modes of accretion onto the main halo is shown to be inherently contained in the merger tree (and would be similarly contained in a statistically generated merger tree which may be used in the study of larger volumes).

The subsequent fate of accreted gas is not modelled equivalently by the two methods, but the analytic framework {\em does} account for both shocked and unshocked gas. Furthermore, the estimated formation time of the hot halo is in excellent agreement with the simulation. 

The single-parameter analytic distributions which are assumed in {\sc galform} are found to be a satisfactory description of the radial distribution of gas and stars inside the simulated halo. It was also shown that the choice of hot halo scalelength is of minor importance in the calculation of cooled gas mass.

However, the enforcement of spherical symmetry in these analytic distributions means that any subsequent cold accretion along filaments (persisting in spite of shocks elsewhere) is neglected. Thus, the {\em total} quantity of unshocked gas onto the central galaxy is underestimated, as expected by \scite{Brooks09}. Some gas is consequently delayed in its arrival at the central galaxy, though it is found to proceed at later times.

The disk structure is predicted analytically by conserving the angular momentum of the cooled gas. Resulting scalelengths are within a factor of two of the simulation results and the circular velocities even closer. This is quite satisfactory for the case of a single halo, but more work is needed to establish whether the two techniques will consistently agree to this level for an arbitrary set of initial conditions (and whether any disagreement is biased).

With regard to the subsequent evolution of the disk, the {\sc galform} code was altered to apply the same fundamental assumptions as the simulation by considering their {\em integrated effect over the assumed disk structure}. Subsequent agreement was excellent given this immense reduction in complexity from many thousands of SPH particles to a few one-dimensional equations.

So, by assuming the same physics and using the same initial conditions, the two techniques predict a final system which is recognisably the same, despite fundamental differences in the way that these assumptions are applied and the evolution followed. This suggests that equations which attempt to characterise the emergent behaviour of the system may, with sufficient understanding, become as reliable as those which emulate the properties of particles themselves.

With such {\em potential} consistency thus established, the same system was then evolved under physical assumptions which have been more frequently adopted within {\sc galform} \cite{Bower06}. The resulting system (at any time) is {\em not} recognisably the same as that predicted by the simulation: having less than a quarter of the stellar mass, barely half the rotational velocity and with negligible star formation at z=0. Though not all aspects of a system need to be correctly represented to learn from the modelling exercise, this divergence from the same initial conditions proves that at least one of these published models poorly represents the true nature of galaxy formation.

It is hoped that this detailed comparison between two very different theoretical approaches will promote the awareness of their respective limitations and help lead them both towards a truer description of real galaxies.

\section*{Acknowledgments}

The authors would primarily like to thank the KITP, Santa Barbara for their hospitality, supported in part by the National Science Foundation under Grant No. PHY05-51164, without which this interdisciplinary study would not have taken place.

AMB acknowledges support from the Sherman Fairchild Foundation. AJB acknowledges support from the Gordon \& Betty Moore Foundation. FG acknowledges support from the HST GO-1125, NSF AST-0607819 and NASA ATP NNX08AG84G grants.

Thanks go to Richard Bower, Shaun Cole and Carlos Frenk for their very helpful comments. Also, they and their collegues Carlton Baugh, John Helly, Cedric Lacey and Rowena Malbon allowed us to use the {\sc galform} semi-analytic model of galaxy formation (\href{http://www.galform.org}{\tt www.galform.org}) in this work, for which we are extremely grateful.

\appendix
\section{Satellites}\label{Satellites}

The orbits of sub halos around their host can be characterised \cite{Lacey93} by the dimensionless parameter
\eq
\Theta = \left(\frac{j}{j_c(E)}\right)^{0.78}\left(\frac{r_{\rm c}(E)}{r_{\rm v}}\right)^2~,\label{ThetaOrbit}
\qe
where $r_c(E)$ and $j_c(E)$ are the radius and specific angular momentum of a circular orbit with energy $E$. For a potential with a flat rotation curve; where the circular velocity at all radii is $v_{\rm c}$,
\eq
\Phi(r) = \Phi(r_{\rm v})  + v_{\rm c}^2\ln\left(\frac{r}{r_{\rm v}}\right)~.
\qe
Hence, for a subhalo at position, ${\bf r}$ with velocity ${\bf v}$, the radius of a circular orbit with the same energy is:
\eq\label{r_c}
r_{\rm c}({\bf v},{\bf r}) = |{\bf r}|\exp\left[ \frac{{\bf v}^2}{2v_{\rm c}^2} - \frac{1}{2} \right]~.
\qe
The orbital parameters are found by simply applying equations (\ref{ThetaOrbit}) to (\ref{r_c}) to the position and velocity vectors of each satellite halo (after moving to a frame where these vectors are both {\bf 0} for the host halo):
\ar
&\Theta&=\left(\frac{j}{v_cr_{\rm v}}\right)^{0.78}\left(\frac{r_{\rm c}(E)}{r_{\rm v}}\right)^{1.22}\nonumber \\
&=& \left(\frac{|{\bf v}\times{\bf r}|}{v_{\rm c}r_{\rm v}}\right)^{0.78}\left(\frac{|{\bf r}|}{r_{\rm v}}\right)^{1.22}\exp\left[0.61\left( \frac{{\bf v}^2}{v_{\rm c}^2}  - 1 \right)\right]~.\label{OrbitFactor}
\ra
The {\sc galform} model assigns each satellite a value for this parameter, $\Theta$, drawn at random from the log-normal distribution that it was found to follow in simulations by \scite{Tormen97}:  
\eq\label{OrbitDist}
\frac{\dif \ln n}{\dif\ln\Theta}=\frac{1}{\sigma\sqrt{2\pi}}\exp\left[-\frac{\left(\ln\Theta+0.14\right)^2}{2\sigma^2}\right]~,
\qe
with $\sigma=0.26$. The sample of orbit parameters produced by this random assignment is shown in Fig. \ref{orbits}, together with the actual values from the \scite{Brooks09} simulation. 

\fig
\includegraphics[trim = 12mm 54mm 10mm 143mm, clip, width=\columnwidth]{orbits}
\caption{{\bf Left panel:} The velocities and radii of the satellite halos in the simulation of \protect\scite{Brooks09}, at the timestep immediately before they are deemed to have merged with their host. The dots are those which appear in the distribution of $\Theta$ in Fig. \ref{Theta}, crosses are those that have $\Theta>4.5$ and hence do not appear. The line shows the constant value of $\Theta$ corresponding to that cut-off, for the given approach angle. {\bf Right panel:} The distribution of approach angles of these satellites alongside the distribution found from the study by \protect\scite{Benson05}.}\label{orbits}
\gif

The fraction of the halo free-fall time that it takes for the satellite to merge through dynamical friction is assumed  to be proportional to $\Theta$:
\eq\label{t_mrg}
\frac{\tau_{\rm mrg}}{\left[G\rho_{\rm v}(z)\right]^{-\frac{1}{2}}} = \Theta\frac{0.3722}{\ln\left(M_{\rm host}/M_{\rm sat}\right)}\frac{M_{\rm host}}{M_{\rm sat}}~.
\qe
This is based on the standard Chandrasekhar formula for the dynamical friction and appears originally in \scite{Lacey93}. $\rho_{\rm v}$ is the mean density at the virial radius (determined by the cosmology, not the specific halo's properties). 

The distribution of angles at which sub structures enter their host halo has been studied by \scite{Benson05} for the VLS and VIRGO simulations and and found to have a repeatable distribution. This distribution can be applied in the {\sc galform} model to generate an alternative set of orbital parameters, $\Theta$, which are shown in Fig. \ref{orbits} alongside the standard assumption of the log-normal distribution, and the results from {\sc gasoline}.

Merger times predicted by (\ref{t_mrg}) have been compared by \scite{Jiang08} with the results of GADGET2 simulations \cite{Springel05}. The agreement found was of the order of a factor of two, which is deemed acceptable for the continued use of this formula in {\sc galform}. However, since (\ref{t_mrg}) was found to consistently underestimate the simulated merger time, the improved fitting formula proposed by\scite{Jiang08} may well be adopted in future.  

Unfortunately, due to the very different definition of merging adopted by the halo finder used here (\S\ref{MergerTree}), a meaningful comparison of the respective times in the two realisation has not been possible.

\section{Disk Profiles}

The stellar disk radii that appear in Fig. \ref{Radii} are generated from analysis of the distribution of stars in the simulation. Fig. \ref{stars} compares the distribution of stellar mass from the simulation with the analytic forms assumed by {\sc galform}, both for disk stars (\ref{RadialProfile}) and for the mass of stars in the bulge, assumed to be distributed such that the projected surface density profile is given by:
\eq
\Sigma_{\rm bulge} \propto \exp\left[-\left(\frac{r}{r_{\rm b}}\right)^\frac{1}{4}\right]~.\label{rquarter}
\qe

Though the agreement is not comprehensive, the analytic forms are an adequate description of the global distribution of simulated stars for the majority of the systems evolution. Unsurprisingly, the simple profiles of (\ref{RadialProfile}) and (\ref{rquarter}) fail to describe the simulated system at early times, before an ordered galactic system has formed.

The fact that {\sc galform} adheres, at {\em all} times, to the fitting forms that are really only appropriate to recent times ($z<2$) is clearly a valid criticism of this technique. However, as long as such forms are indeed a good description of low-redshift systems, then the only way their earlier misjudgements can significantly mislead the predictions of the model is in the early time star formation rates (Fig.\ref{SFtimescale}), which are based on this assumed structure.

\fig
\includegraphics[trim = 7mm 55mm 10mm 50mm, clip, width=\columnwidth]{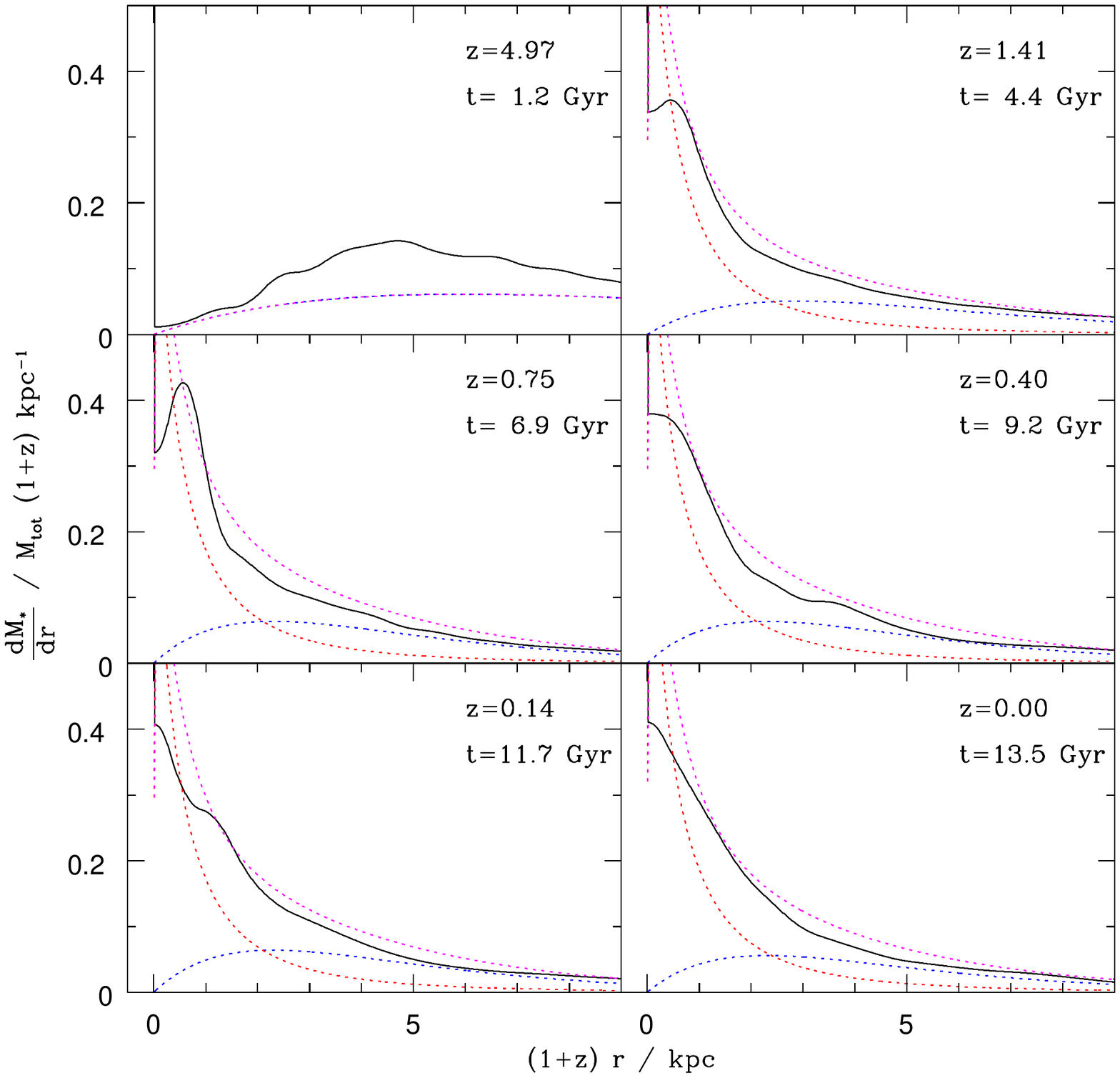}
\caption{The stellar mass profile from the simulation of \protect\scite{Brooks09} at six different redshifts (solid black lines).  The dotted blue lines show the analytic forms which are assumed by the {\sc galform} model for the disk (\ref{RadialProfile}). Dotted red lines show the bulge components (\ref{rquarter}). The total stellar contribution, shown by dotted magenta lines, is constrained to integrate to the same mass as the soild lines. The two lines are therefore fitted using three free parameters: The relative fraction of mass in the two components, the disk scalelength, $r_{\rm D}$ and the scalelength of the bulge profile, $r_{\rm b}$.}\label{stars}
\gif



\begin{thebibliography}{widestentry}

\bibitem[Agertz et al.<2009>]{Agertz09}
Agertz, O., Teyssier, R., \& Moore, B. (2009), MNRAS, 397, 64

\bibitem[Bower et al.<1993>]{Bower93} 
Bower, R.~G., Coles, P., Frenk, C.~S., \& White, S.~D.~M.\ 1993, ApJ, 405, 403 

\bibitem[Bower et al.<2006>]{Bower06} Bower,~R. G., Benson,~A.~J., Malbon,~R., Helly,~ J.~C., Frenk,~C.~S., Baugh,~C.~M., Cole,~S. \& Lacey,~C.~G. 2006, MNRAS, 370, 645

\bibitem[Benson et al.<2001>]{Benson01} 
Benson, A.~J., Pearce, F.~R., Frenk, C.~S., Baugh, C.~M., \& Jenkins, A. 2001, MNRAS, 320, 261 

\bibitem[Benson<2005>]{Benson05} 
Benson,~A.~J. 2005, MNRAS, 358, 551 

\bibitem[Benson et al.<in prep.>]{Benson09}
Benson,~A.~J. et al. (in prep.)

\bibitem[Birnboim \& Dekel<2003>]{Birnboim03} 
Birnboim, Y., \& Dekel, A. 2003, MNRAS, 345, 349 

\bibitem[Blumenthal et al.<1986>]{Blumenthal86}
Blumenthal~G.~R.,~Faber~S.~M., Flores~R. \& Primack~J.~R., 1986, ApJ,  301, 27

\bibitem[Brooks et al.<2007>]{Brooks07} 
Brooks, A.~M., Governato, F., Booth, C.~M., Willman, B., Gardner, J.~P., Wadsley, J., Stinson, G., \& Quinn, T. 2007, ApJ, 655, L17 

\bibitem[Brooks et al.<2009>]{Brooks09}
Brooks, A.~M., Governato, F., Quinn, T., Brook, C.~B.,  \& Wadsley, J. 2009, APJ, 694, 396 


\bibitem[Cattaneo et al.<2007>]{Cattaneo07} 
Cattaneo, A., et al. 2007, MNRAS, 377, 63 

\bibitem[Cole et al.<2000>]{Cole00}
Cole~S., Lacey~C., Baugh~C. \&  Frenk~C. 2000, MNRAS, 319, 168 

\bibitem[Croton et al.<2006>]{Croton06} 
Croton, D.~J., et al.\ 2006, MNRAS, 365, 11 

\bibitem[De Lucia \& Blaizot<2007>]{DeLucia07} 
De Lucia, G., \& Blaizot, J. 2007, MNRAS, 375, 2 

\bibitem[Dekel et al.<2009>]{Dekel09} 
Dekel, A., et al.\ 2009, Nature, 457, 451 

\bibitem[Efstathiou, Lake \& Negroponte<1982>]{Efstathiou82} 
Efstathiou, G., Lake, G., \& Negroponte, J. 1982, MNRAS, 199, 1069 

\bibitem[Gill, Knebe, \& Gibson<2004>]{Knebe04}
Gill,~P.~D.~S., Knebe,~A \& Gibson,~B.~K. 2004, MNRAS, 351, 399 

\bibitem[Governato et al.<2007>]{Governato07}
Governato, F., Willman, B., Mayer, L., Brooks, A., Stinson, G., Valenzuela, O., Wadsley, 
J., \& Quinn, T. 2007, MNRAS, 374, 1479 


\bibitem[Governato, Mayer \& Brook<2008>]{Governato08}
Governato, F., Mayer, L., \& Brook, C. 2008, Astronomical Society of the Pacific Conference Series, 396, 453 

\bibitem[Governato et al.<2009>]{Governato09}
Governato, F., Brook, C. B., Brooks, A. M., Mayer, L., Willman, B., Jonsson, P., Stilp, A. M., Pope, L., Christensen, C., Wadsley, J., Quinn, T.  2008, arXiv:0812.0379 

\bibitem[Gill, Knebe \& Gibson<2004>]{Gill04}
 Gill, S.~P.~D., Knebe, A., \& Gibson, B.~K. 2004, MNRAS, 351, 399 

\bibitem[Gross<1997>]{Gross97}
Gross, M.~A.~K.\ 1997, Ph.D.~Thesis,  

\bibitem[Haardt \& Madau<1996>]{Haardt96} 
Haardt, F., \& Madau, P. 1996, ApJ, 461, 20 

\bibitem[Haardt \& Madau<2001>]{Haardt01} 
Haardt, F., \& Madau, P. 2001, Clusters of Galaxies and the High Redshift Universe Observed in X-rays,  

\bibitem[Hatton et al.<2003>]{Hatton03} 
Hatton, S., Devriendt, J.~E.~G., Ninin, S., Bouchet, F.~R., Guiderdoni, B., \& Vibert, D.\ 2003, MNRAS, 343, 75 

\bibitem[H{\"a}ring \& Rix<2004>]{Haring04} 
H{\"a}ring, N., \& Rix, H.-W. 2004, ApJ, 604, L89 

\bibitem[Helly et al.<2003>]{Helly03} 
Helly, J.~C., Cole, S., Frenk, C.~S., Baugh, C.~M., Benson, A., \& Lacey, C.\ 2003, MNRAS, 338, 903

\bibitem[Jiang et al.<2008>]{Jiang08} Jiang, C.~Y., Jing, 
Y.~P., Faltenbacher, A., Lin, W.~P., \& Li, C. 2008, ApJ, 675, 1095 

\bibitem[Katz \& White<1993>]{Katz93} 
Katz, N., \& White, S.~D.~M. 1993, ApJ, 412, 455 

\bibitem[Kauffmann, White \& Guiderdoni<1993>]{Kauffmann93}
 Kauffmann, G., White, S.~D.~M., \& Guiderdoni, B. 1993, MNRAS, 264, 201 

\bibitem[Kauffmann et al.<1999>]{Kauffmann99} 
Kauffmann, G., Colberg, J.~M., Diaferio, A., \& White, S.~D.~M.\ 1999, MNRAS, 303, 188 

\bibitem[Kere{\v s} et al.<2005>]{Keres05} 
Kere{\v s}, D., Katz, N., Weinberg, D.~H., \& Dav{\'e}, R. 2005, MNRAS, 363, 2 

\bibitem[Knebe, Green \& Binney<2001>]{Knebe01} 
Knebe,~A.~G., Green,~A. \& Binney~J., 2001 MNRAS, 325, 845

\bibitem[Knollmann \& Knebe<2009>]{Knollmann09}
 Knollmann, S.~R., \& Knebe, A. 2009, ApJS, 182, 608 

\bibitem[Kroupa<1993>]{Kroupa93}
Kroupa, P., Tout, C.~A., \& Gilmore, G. 1993, MNRAS, 262, 545 

\bibitem[Kennicutt<1998>]{Kennicutt98} 
Kennicutt~R.~C., 1998, Annual Review of Astronomy \& Astrophysics, 36, 189

\bibitem[Maiolino et al.<2008>]{Maiolino08} Maiolino, R., et al.\ 2008, A\& A, 488, 463 

\bibitem[Ocvirk et al.<2008>]{Ocvirk08} 
Ocvirk, P., Pichon, C., \& Teyssier, R. 2008, MNRAS, 390, 1326 

\bibitem[Pontzen et al.<2008>]{Pontzen08} 
Pontzen, A., et al. 2008, MNRAS, 390, 1349
 
\bibitem[Press \& Schechter<1974>]{Press74} 
Press, W.~H., \& Schechter, P. 1974, ApJ, 187, 425 

\bibitem[Quinn, Katz \& Efstathiou<1996>]{Quinn96}
Quinn, T., Katz, N., \& Efstathiou, G. 1996, MNRAS, 278, L49 


\bibitem[Read et al.<2009>]{Read09}
Read, J.~I., Mayer, L., Brooks, A.~M., Governato, F., \& Lake, G.\ 2009, arXiv:0902.0009 

\bibitem[Sutherland \& Dopita<1993>]{Sutherland93}
Sutherland~R.~S. \& Dopita~M.~A., 1993, ApJ,  88, 253

\bibitem[Lacey \& Cole<1993>]{Lacey93}
Lacey~C. \& Cole~S. 1993, MNRAS, 262, 627

\bibitem[Maiolino et al.<2008>]{Maiolino08}
Maiolino, R., Nagao, T., Grazian, A., Cocchia, F., Marconi, A., Mannucci, F., Cimatti, A., Pipino, A., and 11 co-authors

\bibitem[Mayer, Governato \& Kaufmann<2008>]{Mayer08}
Mayer, L., Governato, F., \& Kaufmann, T. 2008, Advanced Science Letters, 1, 7 

\bibitem[Moore et al.<1999>]{Moore99}
Moore, B., Ghigna, S., Governato, F., Lake, G., Quinn, T., Stadel, J., \& Tozzi, P. 1999, ApJL, 524, L19 

\bibitem[Sharma \& Steinmetz<2005>]{Sharma05}
Sharma~S. \& Steinmetz~M., 2005, ApJ, 628, 21 

\bibitem[Springel<2005>]{Springel05}
Springel~V. 2005, MNRAS, 364, 1105

\bibitem[Stringer \& Benson<2007>]{Stringer07}
Stringer~M.~J. \& Benson,~A.~J. 2007, MNRAS, 382, 641

\bibitem[Stringer et al.<2009>]{Stringer09} 
Stringer, M.~J., Benson, A.~J., Bundy, K., Ellis, R.~S., 
\& Quetin, E.~L.\ 2009, MNRAS, 393, 1127 

\bibitem[Stinson et al.<2006>]{Stinson06} Stinson, G., Seth, A., 
Katz, N., Wadsley, J., Governato, F., \& Quinn, T.\ 2006, MNRAS, 373, 1074 

\bibitem[Tormen<1997>]{Tormen97} 
Tormen,~G., 1997, MNRAS, 290, 411

\bibitem[Wadsley, Stadel \& Quinn<2004>]{Wadsley04}
Wadsley, J.~W., Stadel, J., \& Quinn, T. 2004, New Astronomy, 9, 137 

\bibitem[White \& Frenk<1991>]{White91}
White, S.~D.~M., \& Frenk, C.~S. 1991, ApJ, 379, 52 

\bibitem[Yoshida et al.<2002>]{Yoshida02} 
Yoshida, N., Stoehr, F., Springel, V., \& White, S.~D.~M. 2002, MNRAS, 335, 762 

\bibitem[Zolotov et al.<2009>]{Zolotov09} Zolotov, A., Willman, 
B., Brooks, A.~M., Governato, F., Brook, C.~B., Hogg, D.~W., Quinn, T., 
\& Stinson, G. 2009, arXiv:0904.3333 

\end{thebibliography}
\end{document}